\documentclass[%
preprint,
amsmath,amssymb,
aps,
prb,
]{revtex4-2}

\usepackage{graphicx}
\usepackage{dcolumn}
\usepackage{bm}
\usepackage{mathrsfs}
\renewcommand{\vec}[1]{\boldsymbol{#1}}
\usepackage{tabularx,booktabs,caption}
\newcolumntype{Z}{>{\raggedright}X}

\begin{document}

%\preprint{APS/123-QED}

\title{Calculation of Gilbert damping and magnetic moment of inertia using torque-torque correlation model within \textit{ab initio} Wannier framework}% Force line breaks with \\

\author{Robin Bajaj$^1$, Seung-Cheol Lee$^2$, H. R. Krishnamurthy$^1$}
\author{Satadeep Bhattacharjee$^2$}
\email{s.bhattacharjee@ikst.res.in}
\author{Manish Jain$^1$}
\email{mjain@iisc.ac.in}

	\affiliation{$^1$Centre for Condensed Matter Theory, Department of Physics, Indian Institute of Science, Bangalore 560012, India \\
            $^2$Indo-Korea Science and Technology Center, Bangalore 560065, India \\
            }

\date{\today}% It is always \today, today,

\begin{abstract}
Magnetization dynamics in magnetic materials are well described by the modified semiclassical Landau-Lifshitz-Gilbert (LLG) equation, which includes the magnetic damping $\hat{\bm{\alpha}}$ and the magnetic moment of inertia $\hat{\bm{\mathrm{I}}}$ tensors as key parameters. Both parameters are material-specific and physically represent the time scales of damping of precession and nutation in magnetization dynamics. $\hat{\bm{\alpha}}$ and $\hat{\bm{\mathrm{I}}}$ can be calculated quantum mechanically within the framework of the torque-torque correlation model. The quantities required for the calculation are torque matrix elements, the real and imaginary parts of the Green's function and its derivatives. Here, we calculate these parameters for the elemental magnets such as Fe, Co and Ni in an  \textit{ab initio} framework using density functional theory and Wannier functions. We also propose a method to calculate the torque matrix elements within the Wannier framework. We demonstrate the effectiveness of the method by comparing it with the experiments and the previous \textit{ab initio} and empirical studies and show its potential to improve our understanding of spin dynamics and to facilitate the design of spintronic devices.
\end{abstract}

\maketitle

\section{\label{sec:level1}INTRODUCTION}

In recent years, the study of spin dynamics\cite{fahnle2005fast,antropov1995ab,skubic2008method,antropov1996spin,steiauf2005damping} in magnetic materials has garnered significant attention due to its potential applications in spintronic devices and magnetic storage technologies\cite{parkin2003magnetically,xu2006spintronic,kim2016chiral,chumak2015magnon}. Understanding the behaviour of magnetic moments and their interactions with external perturbations is crucial for the development of efficient and reliable spin-based devices. Among the various parameters characterizing this dynamics, Gilbert damping\cite{gilbert2004phenomenological} and magnetic moment of inertia play pivotal roles. The fundamental semi-classical equation describing the magnetisation dynamics using these two crucial parameters is the Landau-Lifshitz-Gilbert (LLG) equation\cite{bhattacharjee2012atomistic,ciornei2011magnetization}, given by:
\begin{equation}\label{eqn:1}
        \frac{\partial\vec{M}}{\partial t} = \vec{M}\times\left(-\gamma\vec{H} + \frac{\hat{\bm{\alpha}}}{M}\frac{\partial\vec{M}}{\partial t}+\frac{\hat{\bm{\mathrm{I}}}}{M}\frac{{\partial}^{2}\vec{M}}{\partial t^{2}}\right)
\end{equation}
\noindent
where $\bf{M}$ is the magnetisation, $\bf{H}$ is the effective magnetic field including both external and internal fields, $\hat{\bm{\alpha}}$ and $\hat{\bm{\mathrm{I}}}$ are the Gilbert damping and moment of inertia tensors with the tensor components defined as $\alpha^{\mu\nu}$ and $\mathrm{I}^{\mu\nu}$, respectively, and $\gamma$ is the gyromagnetic ratio.
\noindent
Gilbert damping, $\hat{\bm{\alpha}}$ is a fundamental parameter that describes the dissipation of energy during the precession of magnetic moments in response to the external magnetic field. Accurate determination of Gilbert damping is essential for predicting the dynamic behaviour of magnetic materials and optimizing their performance in spintronic devices. On the other hand, the magnetic moment of inertia, $\hat{\bm{\mathrm{I}}}$ represents the resistance of a magnetic moment to changes in its orientation. It governs the response time of magnetic moments to external stimuli and influences their ability to store and transfer information. The moment of inertia\cite{bottcher2012significance} is the magnetic equivalent of the inertia in classical mechanics\cite{wieser2013comparison,chudnovskiy2014spin} and acts as the magnetic inertial mass in the LLG equation.

Experimental investigations of Gilbert damping\cite{fuchs2007spin,oogane2006magnetic,barati2013calculation,bhagat1974temperature,schreiber1995gilbert,inaba2006damping,song2013observation,mizukami2010gilbert} and moment of inertia involve various techniques, such as ferromagnetic resonance (FMR) spectroscopy\cite{trunova2009ferromagnetische,heinrich1966temperature}, spin-torque ferromagnetic resonance (ST-FMR), and time-resolved magneto-optical Kerr effect (TR-MOKE)\cite{zhao2016experimental,li2015inertial}. Interpreting the results obtained from these techniques in terms of the LLG equation provide insights into the dynamical behaviour of magnetic materials and can be used to extract the damping and moment of inertia parameters. In order to explain the experimental observations in terms of more macroscopic theoretical description, various studies\cite{umetsu2012theoretical,thonig2014gilbert,thonig2017magnetic,thonig2018nonlocal,PhysRevLett.99.027204,PhysRevLett.107.066603,gilmore2009evaluating} based on linear response theory and Kambersky theory have been carried out.

Linear response theory-based studies of Gilbert damping and moment of inertia involve perturbing the system and calculating the response of the magnetization to the perturbation. By analyzing the response, one can extract the damping parameter. \textit{Ab initio} calculations based on linear response theory\cite{PhysRevLett.107.066603} can provide valuable insights into the microscopic mechanisms responsible for the damping process. While formal expression for the moment of inertia in terms of Green's functions have been derived within the Linear response framework\cite{bhattacharjee2012atomistic}, to the best of our knowledge, there hasn't been any first principle electronic structure-based calculation for the moment of inertia within this formalism. 

Kambersky's theory\cite{kambersky2007spin,kunevs2002first,kambersky1984fmr} describes the damping phenomena using a breathing Fermi surface~\cite{KB} and torque-torque correlation model~\cite{TT}, wherein the spin-orbit coupling acts as the perturbation and describes the change in the non-equilibrium population of electronic states with the change in the magnetic moment direction. Gilmore \textit{et al.}\cite{PhysRevLett.99.027204,gilmore2009evaluating} have reported the damping for ferromagnets like Fe, Ni and Co using Kambersky's theory in the projector augmented wave method\cite{liechtenstein1987local}.

The damping and magnetic inertia have been derived within the torque-torque correlation model by expanding the effective dissipation field in the first and second-time derivatives of magnetisation\cite{thonig2017magnetic,thonig2014gilbert,thonig2018nonlocal}. In this work, the damping and inertia were calculated using the combination of first-principles fully relativistic multiple scattering Korringa–Kohn–Rostoker (KKR) method and the tight-binding model for parameterisation\cite{papaconstantopoulos2003slater}. However, there hasn't been any full \textit{ab intio} implementation using density functional theory (DFT) and Wannier functions to study these magnetic parameters.

The expressions for the damping and inertia involves integration over crystal momentum $\bm{k}$ in the first Brillouin zone. Accurate evaluation of the integrals involved required a dense $\bm{k}$-point mesh of the order of $10^{6}-10^{8}$ points for obtaining converged values. Calculating these quantities using full \textit{ab initio} DFT is hence time-consuming. To overcome this problem, here we propose an alternative. To begin with, the first principles calculations are done on a coarse $\bm{k}$ mesh instead of dense $\bm{k}$ mesh. We then utilize the maximally localised Wannier functions (MLWFs)\cite{marzari2012maximally} for obtaining the interpolated integrands required for the denser $\bm{k}$ meshes. In this method, the gauge freedom of Bloch wavefunctions is utilised to transform them into a basis of smooth, highly localised Wannier wavefunctions. The required real space quantities like the Hamiltonian and the torque-matrix elements are calculated in the Wannier basis using Fourier transforms. The integrands of integrals can then be interpolated on the fine $\bm{k}$ mesh by an inverse Fourier transform of the maximally localised quantities, thereby enabling the accurate calculations of the damping and inertia.

The rest of this paper is organized as follows: In Sec. \ref{sec:level2}, we introduce the expressions for the damping and the inertia. We describe the formalism to calculate the two key quantities: Green's function and torque matrix elements, using the Wannier interpolation. In Sec. \ref{sec:level3}, we describe the computational details and workflow.  In Sec. \ref{sec:level4}, we discuss the results for ferromagnets like Fe, Co and Ni, and discuss the agreement with the experimental values and the previous studies. In Sec. \ref{sec:level5}, we conclude with a short summary and the future prospects for the methods we have developed.

\section{\label{sec:level2}THEORETICAL FORMALISM}

First, we describe the expressions for Gilbert damping and moment of inertia within the torque-torque correlation model. Then, we provide a brief description of the MLWFs and the corresponding Wannier formalism for the calculation of torque matrix elements and the Green's function.

\subsection{\label{sec:level2a}Gilbert damping and Moment of inertia within torque-torque correlation model}
If we consider the case when there is no external magnetic field, the electronic structure of the system can be described by the Hamiltonian,
\begin{equation}\label{eqn:2}
\mathcal{H}=\mathcal{H}_0+\mathcal{H}_{exc}+\mathcal{H}_{so}=
\mathcal{H}_{sp}+\mathcal{H}_{so}
\end{equation}
The paramagnetic band structure is described by $\mathcal{H}_0$ and $\mathcal{H}_{exc}$ describes the effective local electron-electron interaction, treated within a spin-polarised (sp) local Kohn-Sham exchange-correlation (exc) functional approach, which gives rise to the ferromagnetism. $\mathcal{H}_{so}$ is the spin-orbit Hamiltonian. As we are dealing with ferromagnetic materials only, we can club the first two terms as $\mathcal{H}_{sp}=\mathcal{H}_0+\mathcal{H}_{exc}$.  During magnetization dynamics, (when the magnetization precesses), only the spin-orbit energy of a Bloch state $|\psi_{n\bm{k}}\rangle$ is affected, where $n$ is the band index of the state.
The magnetization precesses around an effective field ${\bf H}_{\text{eff}}={\bf H}_{\text{int}}+{\bf H}_{\text{damp}}+{\bf H}_{\mathrm{I}}$, where ${\bf H}_{\text{int}}$ is the internal field due to the magnetic anisotropy and exchange energies, ${\bf H}_{\text{damp}}$ is the damping field, and ${\bf H}_{\mathrm{I}}$ is the inertial field, respectively. From Eqn. \eqref{eqn:1}, we can see that the damping field ${\bf H}_{\text{damp}}=\frac{\alpha}{M\gamma}\frac{\partial\vec{M}}{\partial t}$, while ${\bf H}_{\mathrm{I}}=\frac{\mathrm{I}}{M\gamma}\frac{{\partial}^{2}\vec{M}}{\partial t^{2}}$. Equating these damping and inertial fields to the effective field corresponding to the change in band energies as magnetization processes, we obtain the mathematical description of the Gilbert damping and inertia. It was proposed by Kambersky~\cite{TT} that the change of the band energies $\frac{\partial \varepsilon_{n\bm{k}}}{\partial \theta^\mu}$ (${\bf\theta}=\theta\hat{n}$ defines the vector for the rotation) can be related to torque operator or matrix depending on how the Hamiltonian is being viewed $\Gamma^{\mu} = [\sigma^{\mu},\mathcal{H}_{\text{so}}]$, where $\sigma^{\mu}$ are the Pauli matrices. Eventually, within the so-called torque-torque correlation model, the Gilbert damping tensor can be expressed as follows:
\begin{equation}\label{eqn:3}
\alpha^{\mu\nu} = \frac{g}{m_{s}\pi}\int\int\left(-\frac{df(\epsilon)}{d\epsilon}\right)Tr [\Gamma^{\mu}(\mathfrak{I}G)(\Gamma^{\nu})^{\dagger}(\mathfrak{I}G)]\frac{d^{3}\vec{k}}{(2\pi)^{3}}d{\epsilon}
\end{equation}
Here trace $Tr$ is over the band indices, and $m_{s}$ is the magnetic moment. Recently, Thonig \text{et al}~\cite{thonig2017magnetic} have extended such an approach to the case of moment of inertia also, where they deduced the moment of inertia tensor components to be,
\begin{eqnarray}\label{eqn:4}
\nonumber
\mathrm{I}^{\nu\mu} &=& \frac{g\hbar}{m_{s}\pi}\int\int f(\epsilon)Tr [ \Gamma^{\nu}(\mathfrak{I}G)(\Gamma^{\mu})^{\dagger}\frac{\partial^{2}}{\partial\epsilon^{2}}(\mathfrak{R}G) \\ &&
+\Gamma^{\nu}\frac{\partial^{2}}{\partial\epsilon^{2}}(\mathfrak{R}G)(\Gamma^{\mu})^{\dagger}(\mathfrak{I}G) ] \frac{d^{3}\vec{k}}{(2\pi)^{3}}d\epsilon
\end{eqnarray}
Here $f(\epsilon)$ is the Fermi function, $(\mathfrak{R}G)$ and $(\mathfrak{I}G)$ are the real and imaginary parts of Green's function $G$ =$(\epsilon+\iota\eta-\mathcal{H})^{-1}$ with $\eta$ as a broadening parameter, $m$ is the magnetization in units of the Bohr magneton, $\Gamma^{\mu} = [\sigma^{\mu},\mathcal{H}_{so}]$ is the $\mu^{th}$ component of torque matrix element operator or matrix, $\mu=x,y,z$. $\alpha$ is a dimensionless parameter, and $\mathrm{I}$ has units of time, usually of the order of femtoseconds.

To obtain the Gilbert damping and moment of inertia tensors from the above two $\bm{k}$-integrals calculated as sums of discrete $\bm{k}$-meshes, we need a large number of  $\bm{k}$-points, such as 
around $10^{6}$, and more than $10^{7}$, for the converged values of $\alpha$ and $\mathrm{I}$ respectively. The reason for the large $\bm{k}$-point sampling in the first Brillouin zone (BZ) is because Green's function becomes more sharply peaked at its poles at the smaller broadening, $\eta$ which need to be used. For $\mathrm{I}$, the number of $\bm{k}$-points needed is more than what is needed for $\alpha$ because $\partial^{2}\mathfrak{R}G/\partial\epsilon^{2}$ has cubic powers in $\mathfrak{R}G$ and/or $\mathfrak{I}G$ as:
    \begin{eqnarray}\label{eqn:5}
    \nonumber
    \frac{\partial^{2}\mathfrak{R}G}{\partial\epsilon^{2}} &=& 2\left[(\mathfrak{R}G)^{3}-\mathfrak{R}G(\mathfrak{I}G)^{2}- \mathfrak{I}G\mathfrak{R}G\mathfrak{I}G -(\mathfrak{I}G)^{2}\mathfrak{R}G \right]
    \\ &&
    \end{eqnarray}

We note that to carry out energy integration in $\alpha$ it is sufficient to consider a limited number of energy points within a narrow range $(\sim k_B T)$ around the Fermi level. This is mainly due to the exponential decay of the derivative of the Fermi function away from the Fermi level. However, the integral for $\mathrm{I}$ involves the Fermi function itself and not its derivative. Consequently, while the Gilbert damping is associated with the Fermi surface, the moment of inertia is associated with the entire Fermi sea. Therefore, in order to adequately capture both aspects, it is necessary to include energy points between the bottom of the valence band and the Fermi level.

\subsection{\label{sec:level2b}Wannier Interpolation}
\subsubsection{\label{sec:level2ba}\textbf{Maximally Localised Wannier Functions (MLWFs)}}
The real-space Wannier functions are written as the Fourier transform of Bloch wavefunctions,
\begin{eqnarray}\label{eqn:6}
    |w_{n\vec{R}}\rangle = \frac{v_{0}}{(2\pi)^{3}}\int_{BZ}d\vec{k}e^{-\iota\vec{k}.\vec{R}}|\psi_{n\vec{k}}\rangle
\end{eqnarray}
where $|\psi_{n\vec{k}}\rangle$ are the Bloch wavefunctions obtained by the diagonalisation of the Hamiltonian at each $\bm{k}$ point using plane-wave DFT calculations. $v_{0}$ is the volume of the unit cell in the real space. 

In general, the Wannier functions obtained by Eqn. \eqref{eqn:6} are not localised. Usually, the Fourier transforms of smooth functions result in localised functions. But there exists a phase arbitrariness of $e^{\iota\phi_{n\bm{k}}}$ in the Bloch functions because of independent diagonalisation at each $\bm{k}$, which messes up the localisation of the Wannier functions in real space.

To mitigate this problem, we use the Marzari-Vanderbilt (MV) localisation procedure\cite{marzari2012maximally,pizzi2020wannier90,PhysRevB.65.035109} to construct the MLWFs, which are given by,
\begin{eqnarray}\label{eqn:7}
    |w_{n\vec{R}}\rangle = \frac{1}{N}\sum_{\vec{q}}\sum_{m=1}^{\mathcal{N}_{\vec{q}}}e^{-\iota\vec{q}.\vec{R}}\mathcal{U}_{mn}^{\vec{q}}|\psi_{m\vec{q}}\rangle
\end{eqnarray}
where $\mathcal{U}_{mn}^{\vec{q}}$ is a $(\mathcal{N}_{\vec{q}}\times \mathcal{N})$ dimensional matrix chosen by disentanglement and Wannierisation procedure. $\mathcal{N}$ are the number of target Wannier functions, and $\mathcal{N}_{\vec{q}}$ are the original Bloch states at each $\bm{q}$ on the coarse mesh, from which $\mathcal{N}$ smooth  Bloch states on the fine $\bm{k}$-mesh are extracted requiring $\mathcal{N}_{\vec{q}}>\mathcal{N}$ for all $\bm{q}$, $N$ is the number of uniformly distributed $\vec{q}$ points in the BZ. The interpolated wavefunctions on a dense $\bm{k}$-mesh, therefore, are given via inverse Fourier transform as:
\begin{eqnarray}\label{eqn:8}
    |\psi_{n\vec{k}}\rangle = \sum_{\vec{R}}e^{\iota\vec{k}.\vec{R}}|w_{n\vec{R}}\rangle
\end{eqnarray}
Throughout the manuscript, we use $\bm{q}$ and $\bm{k}$ for coarse and fine meshes in the BZ, respectively.
\begin{figure}[t]
\includegraphics[width=0.35\textwidth]{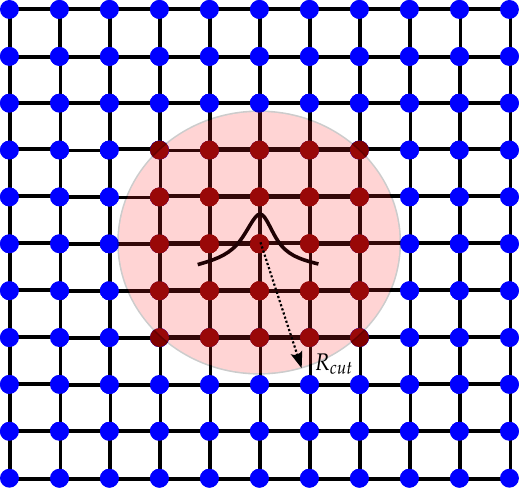}% Here is how to import EPS art
\caption{\label{fig:fft} The figure shows the schematic of the localisation of the Wannier functions on a $\bf{R}$ grid. The matrix elements of the quantities like Hamiltonian on the $\bf{R}$ grid are exponentially decaying. Therefore, most elements on the $\bf{R}$ grid are zero (shown in blue). We can hence do the summation till a cutoff ${R}_{cut}$ (shown in red) to interpolate the quantities on a fine $\bm{k}$ grid.}
\end{figure}

\subsubsection{\label{sec:level2bb}\textbf{Torque Matrix elements}}
As described in the expressions of $\alpha^{\mu\nu}$ and $\mathrm{I}^{\nu\mu}$ in Eqns. \eqref{eqn:3} and \eqref{eqn:4}, the $\mu^{th}$ component of the torque matrix is given by the commutator of $\mu^{th}$ component of Pauli matrices and spin-orbit coupling matrix \textit{i.e.} $\Gamma^{\mu} = [\sigma^{\mu},\mathcal{H}_{so}]$. Physically, we define the spin-orbit coupling (SOC) and spin-orbit torque (SOT) as the dot and cross products of orbital angular momentum and spin angular momentum operator, respectively, such that $\mathcal{H}_{so} = \xi\vec{\ell}.\vec{\mathscr{s}}$ where $\xi$ is the coupling amplitude. Using this definition of $\mathcal{H}_{so}$, one can show easily that $-\iota[\vec{\sigma},\mathcal{H}_{so}] = 2\xi\vec{\ell}\times\vec{\mathscr{s}}$ which represents the torque. 

There have been several studies on how to calculate the spin-orbit coupling using \textit{ab initio} numerical approach. Shubhayan \textit{et al.}\cite{roychoudhury2017spin} describe the method to obtain SOC matrix elements in the Wannier basis calculated without SO interaction, using an approximation of weak SOC in the organic semiconductors considered in their work. Their method involves DFT in the atomic orbital basis, wherein the SOC in the Bloch basis can be related to the SOC in the atomic basis. Then, by the basis transformation, they get the SOC in the Wannier basis calculated in the absence of SO interaction. Farzad \textit{et al.}\cite{PhysRevB.96.214421} calculate the SOC by extracting the coupling amplitude from the Hamiltonian in the Wannier basis, treating the Wannier functions as atomic-like orbitals.

We present a different approach wherein we can do the DFT calculation in any basis (plane wave or atomic orbital). Unlike the previous approaches, we perform two DFT calculations and two Wannierisations: one is with spin-orbit interaction and finite magnetisation (SO) and the other is spin-polarised without spin-orbit coupling (SP). The spin-orbit Hamiltonian, $\mathcal{H}_{so}$ can then be obtained by subtracting the spin-polarized Hamiltonian, $\mathcal{H}_{sp}$ from the full Hamiltonian, $\mathcal{H}$ as $\mathcal{H}_{so} = \mathcal{H} - \mathcal{H}_{sp}$. This, however, can only be done if both the Hamiltonians, $\mathcal{H}$ and $\mathcal{H}_{sp}$, are written in the same basis. We choose to use the corresponding Wannier functions as a basis. It should however be noted that, when one Wannierises the SO and SP wavefunctions, one will get two different Wannier bases. As a result, we can not directly subtract the $\mathcal{H}$ and $\mathcal{H}_{sp}$ in these close but different Wannier bases. In order to do the subtraction, we find the transformation between two Wannier bases {\em, i.e.} express one set of Wannier functions in terms of the other. Subsequently, we can express the matrix elements of $\mathcal{H}$ and $\mathcal{H}_{sp}$ in the same basis and hence calculate $\mathcal{H}_{so}$. In the equations below, the Wannier functions, the Bloch wavefunctions and the operators defined in the corresponding bases in SP and SO calculations are represented with and without the tilde ($\sim$) symbol, respectively.

The $\mathcal{N}$ SO Wannier functions are given by:
\begin{eqnarray}\label{eqn:9}
    |w_{n\vec{R}}\rangle = \frac{1}{N}\sum_{\vec{q}}\sum_{m=1}^{\mathcal{N}_{\vec{q}}}e^{-\iota\vec{q}.\vec{R}}\mathcal{U}_{mn}^{\vec{q}}|\psi_{m\vec{q}}\rangle
\end{eqnarray}
where $\mathcal{U}_{mn}^{\vec{q}}$ is a $(\mathcal{N}_{\vec{q}}\times \mathcal{N})$ dimensional matrix. The wavefunctions and Wannier functions from the SO calculation for a particular $\vec{q}$  and $\vec{R}$ are a mixture of up and down spin states and are represented as spinors:
    \begin{eqnarray}\label{eqn:10}
       |\psi_{n\vec{q}}\rangle =  \begin{bmatrix}
        |\psi_{n\vec{q}}^{\uparrow}\rangle \\
        |\psi_{n\vec{q}}^{\downarrow}\rangle \\
       \end{bmatrix}
        \ \ \ \ \ \
       |w_{n\vec{R}}\rangle =  \begin{bmatrix}
        |w_{n\vec{R}}^{\uparrow}\rangle \\
        |w_{n\vec{R}}^{\downarrow}\rangle \\
       \end{bmatrix}
    \end{eqnarray}

\begin{figure}[h!]
\includegraphics{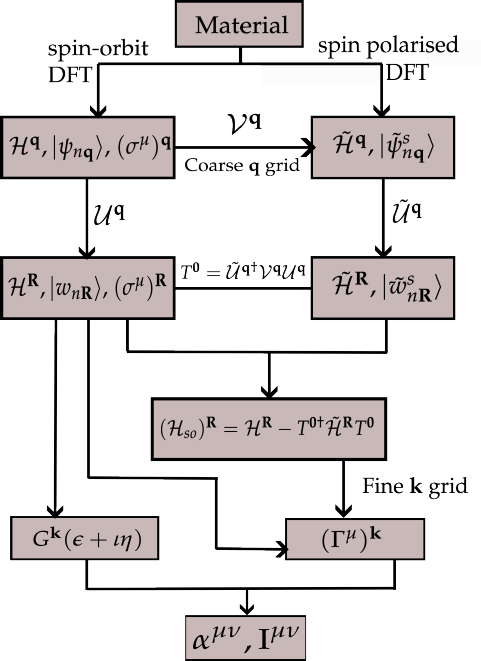}% Here is how to import EPS art
\caption{\label{fig:flowchart} This figure shows the implementation flow chart of the theoretical formalism described in Sec. \ref{sec:level2} }
\end{figure}
The $\Tilde{\mathcal{N}}^{s}$ SP Wannier functions are given by:
\begin{eqnarray}\label{eqn:11}
    |\Tilde{w}_{n\vec{R}}^{s}\rangle = \frac{1}{N}\sum_{\vec{q}}\sum_{m=1}^{\Tilde{\mathcal{N}}_{\vec{q}}^{s}}e^{-\iota\vec{q}.\vec{R}}\Tilde{\mathcal{U}}_{mn}^{\vec{q}s}|\Tilde{\psi}_{m\vec{q}}^{s}\rangle
\end{eqnarray}
where $s=\uparrow,\downarrow$. $\Tilde{\mathcal{U}}_{mn}^{\vec{q}s}$ is a $(\Tilde{\mathcal{N}}_{\vec{q}}^{s}\times \Tilde{\mathcal{N}}^{s})$ dimensional matrix. Since the spinor Hamiltonian doesn't have off-diagonal terms corresponding to opposite spins in the absence of SOC, the wavefunctions will be $|\Tilde{\psi}_{n\vec{q}}^{s}\rangle = |\Tilde{\psi}_{n\vec{q}}\rangle\otimes|s\rangle$. The combined expression for $\Tilde{\mathcal{U}}^{\vec{q}}$ for $\Tilde{\mathcal{N}}^{\uparrow}+\Tilde{\mathcal{N}}^{\downarrow}=\Tilde{\mathcal{N}}$ SP Wannier functions is:
    \begin{eqnarray}\label{eqn:12}
       \Tilde{\mathcal{U}}^{\vec{q}} = \begin{bmatrix}
           \Tilde{\mathcal{U}}^{\vec{q}\uparrow} & 0 \\
           0 & \Tilde{\mathcal{U}}^{\vec{q}\downarrow}
       \end{bmatrix}
    \end{eqnarray}

where $\Tilde{\mathcal{U}}^{\vec{q}}$ is $\Tilde{\mathcal{N}}_{\vec{q}}\times\Tilde{\mathcal{N}}$ dimensional matrix with $\Tilde{\mathcal{N}}_{\vec{q}} = \Tilde{\mathcal{N}}_{\vec{q}}^{\uparrow}+\Tilde{\mathcal{N}}_{\vec{q}}^{\downarrow}$. Dropping the $s$ index for SP kets results in the following expression for $\Tilde{\mathcal{N}}$ SP Wannier functions:
\begin{eqnarray}\label{eqn:13}
    |\Tilde{w}_{n\vec{R}}\rangle = \frac{1}{N}\sum_{\vec{q}}\sum_{m=1}^{\Tilde{\mathcal{N}}_{\vec{q}}}e^{-\iota\vec{q}.\vec{R}}\Tilde{\mathcal{U}}_{mn}^{\vec{q}}|\Tilde{\psi}_{m\vec{q}}\rangle
\end{eqnarray}

We now define the matrix of the transformation between SO and SP Wannier bases as:
\begin{align}\label{eqn:14}
    T_{mn}^{\vec{RR'}} &= \langle \Tilde{w}_{m\vec{R}}|w_{n\vec{R'}}\rangle 
    \nonumber
    \\ & =\frac{1}{N^2}\sum_{\vec{qq'}}\sum_{p,l=1}^{\Tilde{\mathcal{N}}_{\vec{q}},{\mathcal{N}}_{\vec{q'}}}e^{\iota(\vec{q}.\vec{R}-\vec{q'}.\vec{R'})}\Tilde{\mathcal{U}}_{mp}^{\vec{q}\dagger}\langle\Tilde{\psi}_{p\vec{q}}|\psi_{l\vec{q'}}\rangle \mathcal{U}_{ln}^{\vec{q'}}
    \nonumber
    \\ &
    =\frac{1}{N^2}\sum_{\vec{qq'}}e^{\iota(\vec{q}.\vec{R}-\vec{q'}.\vec{R'})}[\Tilde{\mathcal{U}}^{\vec{q}\dagger}\mathcal{V}^{\vec{qq'}}\mathcal{U}^{\vec{q'}}]_{mn}   
\end{align}

Here $\mathcal{V}_{pl}^{\vec{qq'}}=\langle\Tilde{\psi}_{p\vec{q}}|\psi_{l\vec{q'}}\rangle$. Eqn. \eqref{eqn:14} is the most general expression to get the transformation matrix. We can reduce this quantity to a much simpler one using the orthogonality of wavefunctions of different $\vec{q}$. Eqn. \eqref{eqn:14} hence  becomes,
\begin{align}\label{eqn:15}
    T_{mn}^{\vec{RR'}} & =\frac{1}{N^2}\sum_{\vec{q}}e^{\iota\vec{q}.(\vec{R}-\vec{R'})}[\Tilde{\mathcal{U}}^{\vec{q}\dagger}(N\mathcal{V}^{\vec{q}})\mathcal{U}^{\vec{q}}]_{mn}  
    \nonumber
    \\ &
    = \frac{1}{N}\sum_{\vec{q}}e^{\iota\vec{q}.(\vec{R}-\vec{R'})}[\Tilde{\mathcal{U}}^{\vec{q}\dagger}\mathcal{V}^{\vec{q}}\mathcal{U}^{\vec{q}}]_{mn}  
\end{align}
where $\mathcal{V}_{pl}^{\vec{q}}= \langle\Tilde{\psi}_{p\vec{q}}|\psi_{l\vec{q}}\rangle$. Using this transformation, we write SP Hamiltonian in SO Wannier bases as:
\begin{align}\label{eqn:16}
    (\mathcal{H}_{sp})_{mn}^{\vec{RR'}} & = \langle w_{m\vec{R}}|\mathcal{H}_{sp}|w_{n\vec{R'}}\rangle 
    \nonumber
    \\ & = \sum_{pl\vec{R''R'''}}\langle w_{m\vec{R}}|\Tilde{w}_{p\vec{R''}}\rangle
    \nonumber
    \\ & \langle\Tilde{w}_{p\vec{R''}}|\mathcal{H}_{sp}|\Tilde{w}_{l\vec{R'''}}\rangle\langle\Tilde{w}_{l\vec{R'''}}|w_{n\vec{R'}}\rangle
    \nonumber
    \\ &
    = \sum_{pl\vec{R''R'''}}(T^{\dagger})_{mp}^{\vec{RR''}}(\Tilde{\mathcal{H}}_{sp})_{pl}^{\vec{R''R'''}}T_{ln}^{\vec{R'''R'}}
\end{align}
Since Wannier functions are maximally localised and generally atomic-like, the major contribution to the overlap $T_{mn}^{\vec{RR'}}$ is for $\vec{R=R'}$. Therefore, we can write $T_{mn}^{\vec{RR}} = T_{mn}^{\vec{0}}$. The reason is that it depends on relative $\vec{R-R'}$, we can just consider overlaps at $\vec{R}=\vec{0}$. Eqn. \ref{eqn:16} becomes,
\begin{align}\label{eqn:17}
    (\mathcal{H}_{sp})_{mn}^{\vec{RR'}} & = \sum_{pl}(T^{\dagger})_{mp}^{\vec{0}}(\Tilde{\mathcal{H}}_{sp})_{pl}^{\vec{RR'}}T_{ln}^{\vec{0}}
\end{align}
Therefore, we write the $\mathcal{H}_{so}$ in Wannier basis as,
\begin{eqnarray}\label{eqn:18}
    (\mathcal{H}_{so})_{mn}^{\vec{RR'}} = \mathcal{H}_{mn}^{\vec{RR'}}-(\mathcal{H}_{sp})_{mn}^{\vec{RR'}}
\end{eqnarray}
The torque matrix elements in SO Wannier bases are given by,
\begin{eqnarray}\label{eqn:19}
    (\Gamma^{\mu})_{mn}^{\vec{RR'}} = (\sigma^{\mu}\mathcal{H}_{so})_{mn}^{\vec{RR'}} - (\mathcal{H}_{so}\sigma^{\mu})_{mn}^{\vec{RR'}}
\end{eqnarray}
Consider $(\sigma^{\mu}\mathcal{H}_{so})_{mn}^{\vec{RR'}}$ and insert the completeness relation of the  Wannier functions, and also neglecting SO matrix elements between the Wannier functions at different sites because of their being atomic-like.  
\begin{eqnarray}\label{eqn:20}
    (\sigma^{\mu}\mathcal{H}_{so})_{mn}^{\vec{RR'}} & =  \sum_{p\vec{R''}}(\sigma^{\mu})_{mp}^{\vec{RR''}}(\mathcal{H}_{so})_{pn}^{\vec{R''R'}}
    \nonumber
    \\ &
    =  (\sigma^{\mu})_{mp}^{\vec{RR'}}(\mathcal{H}_{so})_{pn}^{\vec{0}}
\end{eqnarray}
$(\sigma^{\mu})_{mp}^{\vec{RR'}}$ is calculated by the Fourier transform of the spin operator written in the Bloch basis, just like the Hamiltonian. 
\begin{eqnarray}\label{eqn:21}
     (\sigma^{\mu})_{mp}^{\vec{RR'}}= \frac{1}{N}\sum_{\vec{q}}e^{-\iota\vec{q}.(\vec{R'-R})}\left[\mathcal{U}^{\vec{q}\dagger}(\sigma^{\mu})^{\vec{q}}\mathcal{U}^{\vec{q}}\right]_{mp}
\end{eqnarray}
We interpolate the SOT matrix elements on a fine $\vec{k}$-mesh as follows:
\begin{eqnarray}\label{eqn:22}
     (\Gamma^{\mu})_{mn}^{\vec{k}}= \sum_{\vec{R'-R}}e^{\iota\vec{k}.(\vec{R'-R})}(\sigma^{\mu})_{mn}^{\vec{RR'}}
\end{eqnarray}

This yields the torque matrix elements in the Wannier basis. In the subsequent expressions, $W$ and $H$ subscripts represent the Wannier and Hamiltonian basis, respectively. In order to rotate to the Hamiltonian gauge, which diagonalises the Hamiltonian interpolated on the fine $\vec{k}$ mesh using its matrix elements in the Wannier basis.
\begin{eqnarray}\label{eqn:23}
     (\mathcal{H}_{W})_{mn}^{\vec{k}}= \sum_{\vec{R'-R}}e^{\iota\vec{k}.(\vec{R'-R})}\mathcal{H}_{mn}^{\vec{RR'}}
\end{eqnarray}
\begin{eqnarray}\label{eqn:24}
     (\mathcal{H}_{H})_{mn}^{\vec{k}}=  \left[(U^{\vec{k}})^{\dagger}(\mathcal{H}_{W})^{\vec{k}}U^{\vec{k}}\right]_{mn}
\end{eqnarray}
Here $U^{\vec{k}}$ (not to be confused with $\mathcal{U}^{\bm{q}}$) are the matrices with columns as the eigenvectors of $(\mathcal{H}_{W})^{\vec{k}}$, and $(\mathcal{H}_{H})_{mn}^{\vec{k}} = \epsilon_{m\bm{k}}\delta_{mn}$. We use these matrices to rotate the SOT matrix elements in Eqn. \eqref{eqn:22} to the Hamiltonian basis as:
\begin{eqnarray}\label{eqn:25}
     (\Gamma_{H}^{\mu})_{mn}^{\vec{k}}=  \left[(U^{\vec{k}})^{\dagger}(\Gamma_{W}^{\mu})^{\vec{k}}U^{\vec{k}}\right]_{mn}
\end{eqnarray}

\subsubsection{\label{sec:level2bc}\textbf{Green's functions}}
The Green's function at an arbitrary $\vec{k}$ and $\epsilon$ on a fine $\vec{k}$-mesh in the Hamiltonian basis is given by:
\begin{eqnarray}\label{eqn:26}
      G_{H}^{\vec{k}}(\epsilon+\iota\eta) = (\epsilon+\iota\eta-(\mathcal{H}_{H})^{\vec{k}})^{-1}
\end{eqnarray}
where $\eta$ is a broadening factor and is caused by electron-phonon coupling and is generally of the order $5-10$ meV. $G_{H}^{\vec{k}}(\epsilon+\iota\eta)$ is a $\mathcal{N}\times\mathcal{N}$ dimensional matrix. 

Therefore, we can calculate $\mathfrak{R}G$, $\mathfrak{I}G$ and $\partial^{2}\mathfrak{R}G/\partial\epsilon^{2}$ as defined in Eqn. \eqref{eqn:5} and hence, $\alpha$ and $\mathrm{I}$.

\section{\label{sec:level3}COMPUTATIONAL DETAILS}
Plane-wave pseudopotential calculations were carried out for the bulk ferromagnetic transition metals bcc Fe, hcp Co and fcc Ni using {\sc Quantum Espresso} package\cite{giannozzi2009quantum,giannozzi2017advanced}. The conventional unit cell lattice constants ($a$) used for bcc Fe and fcc Ni were 5.424 and 6.670 bohrs, respectively and for hcp Co, $a$=4.738 bohr and $c/a$=1.623 were used. The non-collinear spin-orbit and spin-polarised calculations were performed using fully relativistic norm-conserving pseudopotentials. The kinetic energy cutoff was set to 80 Ry. Exchange-correlation effects were treated within the PBE-GGA approximation. The self-consistent calculations were carried out on 16$\times$16$\times$16 Monkhorst-Pack Grid using Fermi smearing of 0.02 Ry. Non-self-consistent calculations were carried out using the calculated charge densities on $\Gamma$-centered 10$\times$10$\times$10 coarse $ \vec{k}$-point grid. For bcc Fe and fcc Ni, 64 bands were calculated and hcp Co 96 bands were calculated (because there are two atoms per unit cell for Co). We define a set of 18 trial orbitals $sp^{3}d^{2}$, $d_{xy}$, $d_{xz}$, and $d_{yz}$ for Fe, 18 orbitals per atom $s$, $p$ and $d$ for Co and Ni, to generate 18 disentangled spinor maximally-localized Wannier functions per atom using {\sc Wannier90} package \cite{pizzi2020wannier90}. 

From the {\sc Wannier90} calculations, we get the checkpoint file \textit{.chk}, which contains all the information about disentanglement and gauge matrices. We use \textit{.spn} and \textit{.eig} files generated by {\sc pw2wannier90} to get the spin operator and the  Hamiltonian in the Wannier basis. We evaluate the SOT matrix elements in the Wannier Basis. 

We get $\alpha$ by simply summing up on a  fine-$\vec{k}$ grid with appropriate weights for the $\vec{k}$-integration, and we use the trapezoidal rule in the range [-8$\delta$,8$\delta$] for energy integration around the Fermi level where $\delta$ is the width of the derivative of Fermi function $\sim k_{B}T$. We consider 34 energy points in this energy range. We perform the calculation for $T=300$K. 

For the calculation of $\mathrm{I}$, we use a very fine grid of $400\times400\times400$ $\vec{k}$-points. For $\eta>0.1$, we use 320 energy points between VBM and Fermi energy. For $0.01<\eta<0.1$, we use 3200-6400 energy points for the energy integration.

\begin{figure*}[t]
\includegraphics[width=\textwidth]{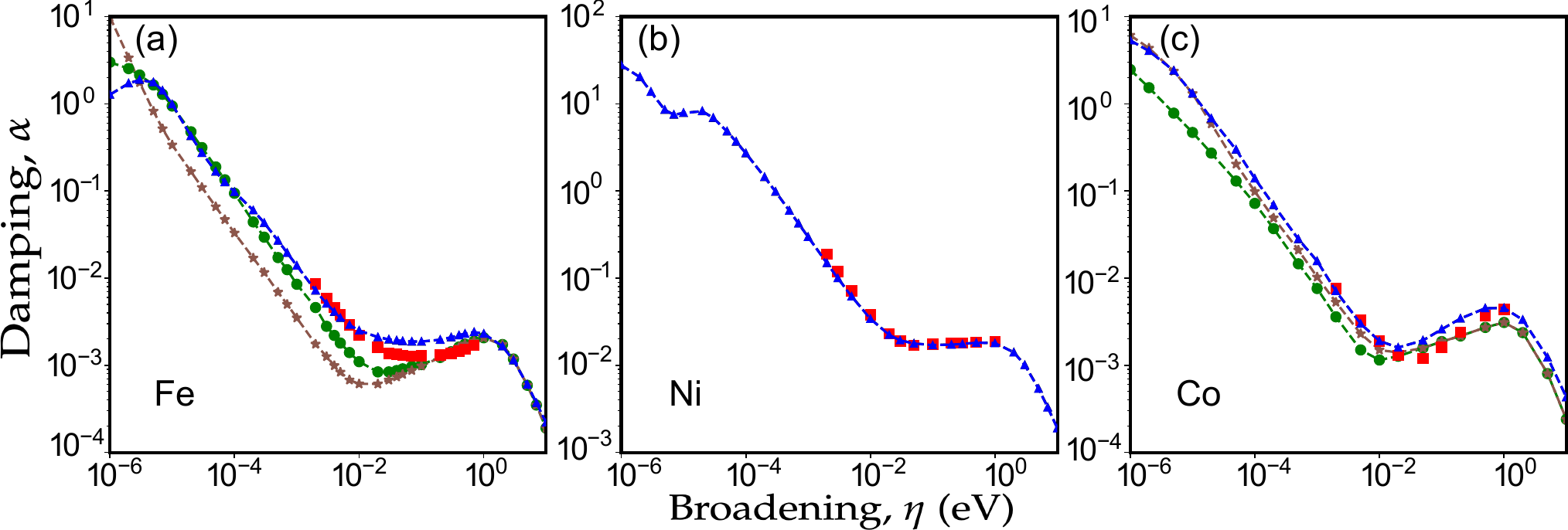}% Here is how to import EPS art
\caption{\label{fig:3} (a), (b) and (c) shows the $\alpha$ vs $\eta$ for Fe, Ni and Co, respectively. Damping constants calculated using the Wannier implementation are shown in blue. Damping calculated using the tight binding method based on Lorentzian broadening and Green's function by Thonig \textit{et al}\cite{thonig2014gilbert} are shown in brown and green, respectively. Comparison with damping constants calculated by Gilmore \textit{et al}\cite{PhysRevLett.99.027204} using local spin density approximation (LSDA) are shown in red. The dotted lines are guides to the eye.}
\end{figure*}
\begin{figure}[h!]
\includegraphics[width=0.43\textwidth]{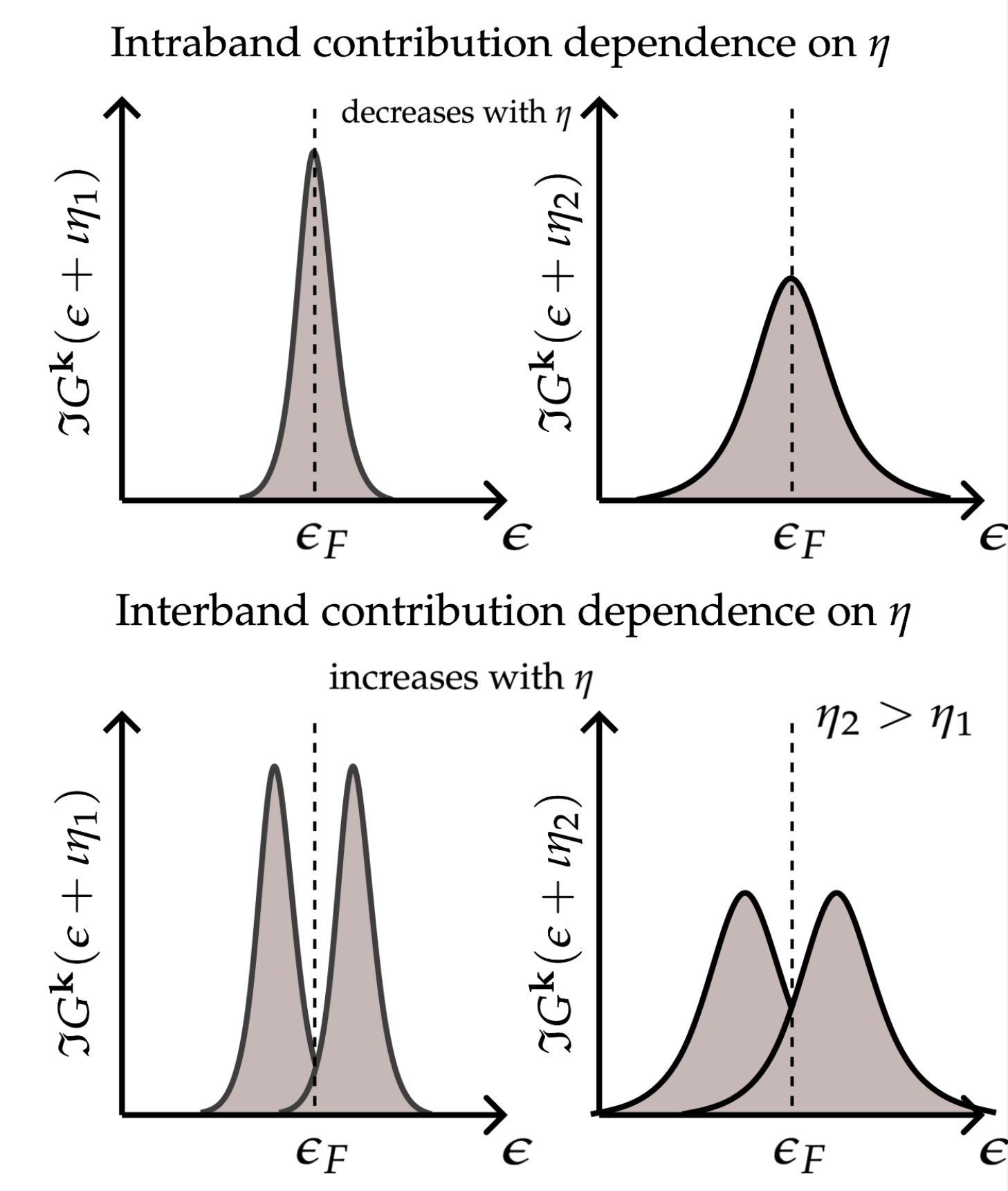}% Here is how to import EPS art
\caption{\label{fig:4} Schematic explaining the dependence of intraband and interband contribution in $\alpha$ with $\eta$.}
\end{figure}

\begin{table}[h!]
\caption{}\label{tab:table}
\setlength\tabcolsep{12pt} % default value: 6pt
\begin{tabularx}{\columnwidth}{@{} Z *{4}{c} @{}}
\toprule
\textrm{Material} & \textrm{$\eta$ (meV)} &
\textrm{-$\mathrm{I}$ (fs)} &
\textrm{$\alpha(\times 10^{-3})$} & \textrm{$\tau$ (ps)} \\
\midrule
Fe & 6 & 0.210 & 3.14 & 0.42 \\
   & 8 & 0.114 & 2.77 & 0.26 \\
   & 10 & 0.069 & 2.51 & 0.17 \\
\midrule
Ni & 10 & 2.655 & 34.2 & 0.48 \\
\midrule
Co & 10 & 0.061 & 1.9 & 0.21\\
\bottomrule 
\end{tabularx}
\end{table}

\section{\label{sec:level4}RESULTS AND DISCUSSION}
\subsection{Damping constant}
In this section, we report the damping constants calculated for the bulk iron, cobalt and nickel. The magnetic moments are oriented in the z-direction. For reference direction z, the damping tensor is diagonal resulting in the effective damping constant $\alpha = \alpha^{xx}+\alpha^{yy}$.

\begin{figure}[b]
\includegraphics[width=0.48\textwidth]{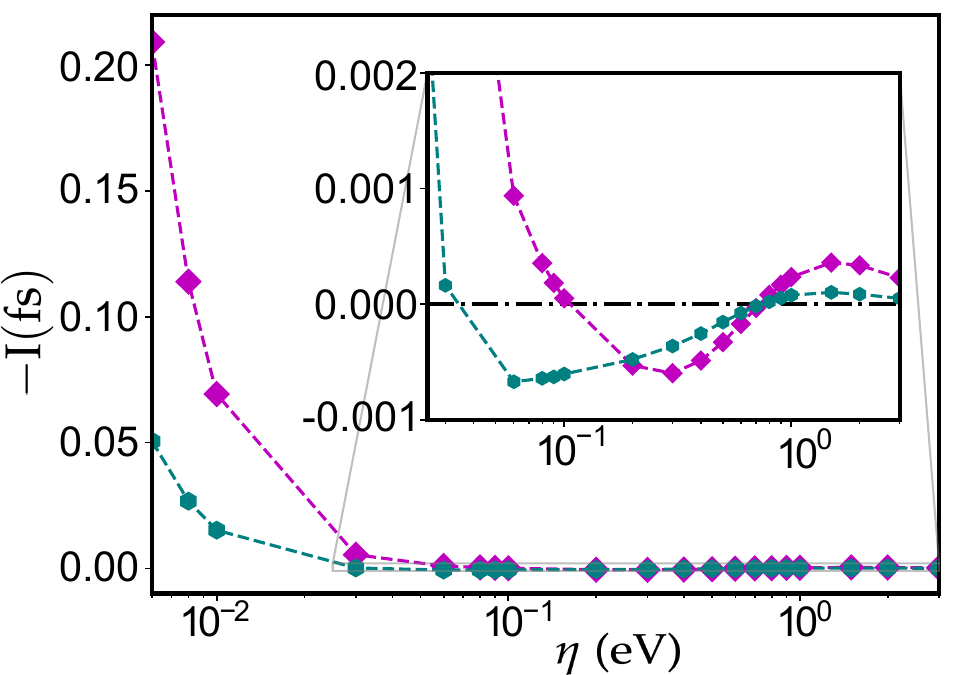}% Here is how to import EPS art
\caption{\label{fig:5} Plot showing moment of inertia, $-\mathrm{I}$ versus broadening, $\eta$. The moment of inertia in the range $0.03-3.0$ eV is shown as an inset. The values using the Wannier implementation and the tight binding method\cite{thonig2017magnetic} are shown in blue and green, respectively.}
\end{figure}

In Fig. \ref{fig:3}, we report the damping constants calculated by the Wannier implementation as a function of broadening, $\eta$ known to be caused by electron-phonon scattering and scattering with impurities. We consider the values of $\eta$ ranging from $10^{-6}$ to $2$ eV to understand the role of intraband and interband transitions as reported in the previous studies\cite{thonig2014gilbert, PhysRevLett.99.027204}. We note that the experimental range is for the broadening is expected to be much smaller with $\eta\sim5-10$ meV. The results are found to be in very good agreement with the ones calculated using local spin density approximation (LSDA)\cite{PhysRevLett.99.027204} and tight binding paramterisation\cite{thonig2014gilbert}.

\begin{figure*}[t]
\includegraphics[width=\textwidth]{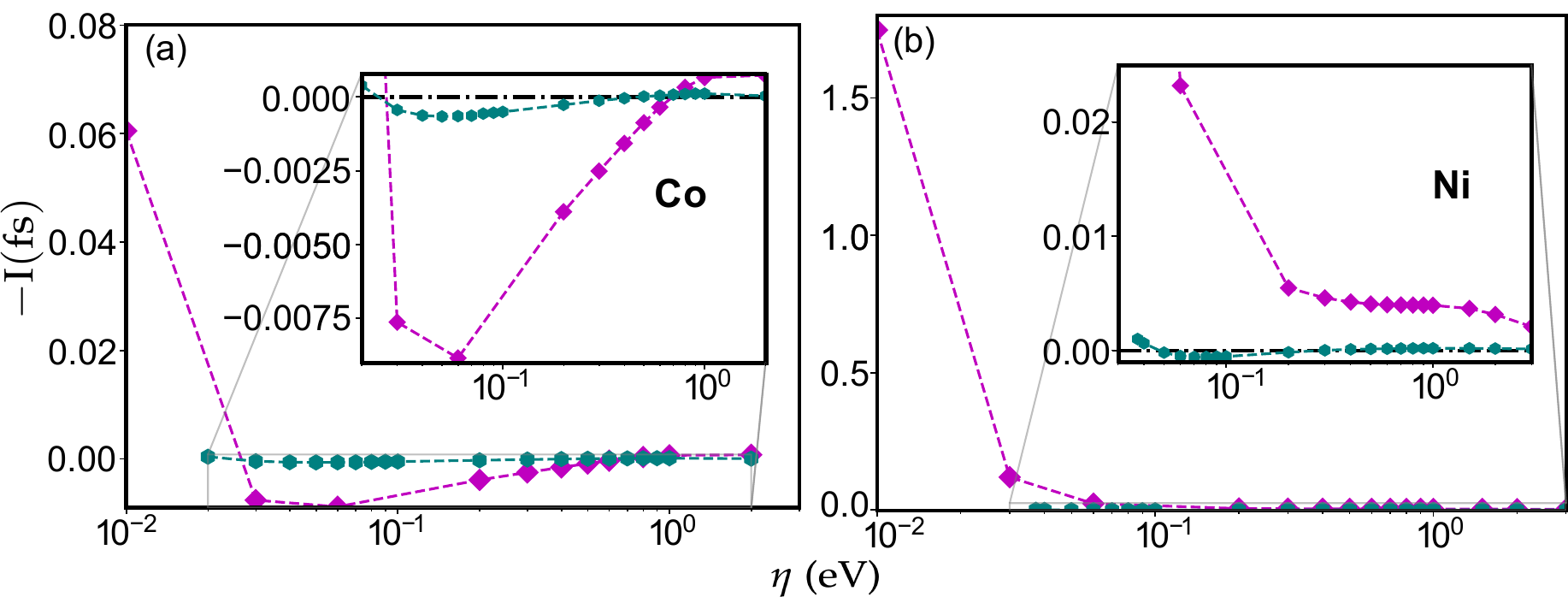}% eHere is how to import EPS art
\caption{\label{fig:6}(a) and (b) show negative of the moment of inertia, $-\mathrm{I}$ versus broadening, $\eta$ for Co and Ni, respectively. The values using the Wannier implementation and the tight binding method are shown in magenta and cyan, respectively. The moment of inertia is shown as an inset in the range $0.03-2.0$ eV and $0.03-3.0$ eV for Co and Ni, respectively. }
\end{figure*}

The expression for Gilbert damping[\ref{eqn:3}] is written in terms of the imaginary part of Green's functions. Using the spectral representation of Green's function, $A_{n\boldsymbol{k}}(\omega)$, we can rewrite Eqn. (\ref{eqn:3}) as:
\begin{eqnarray}\label{eqn:27}
{\alpha}^{\mu \nu }=\frac{g\pi }{m_{s}}\sum _{nm}\int {T}_{nm}^{\mu }({\boldsymbol{k}}){T}_{nm}^{\ast \nu }({\boldsymbol{k}}){S}_{nm}{\rm{d}}{\boldsymbol{k}}
\end{eqnarray}
where ${S}_{nm}=\int \eta (\epsilon ){A}_{n{\boldsymbol{k}}}(\epsilon ){A}_{m{\boldsymbol{k}}}(\epsilon ){\rm{d}}\epsilon$ is the spectral overlap.
Although we are working in the basis where the Hamiltonian is diagonal, the non-zero off-diagonal elements in the torque matrix lead to both intraband ($m=n$) and interband ($m\neq n$) contributions. For the sake of simple physical understanding, we consider the contribution of the spectral overlaps at the Fermi level for both intraband and interband transitions in Fig. \ref{fig:4}. But in the numerical calculation temperature broadening has also been considered. For the smaller $\eta$, the contribution of intraband transitions decreases almost linearly with the increase in $\eta$ because the overlaps become less peaked. Above a certain $\eta$, the interband transitions become dominant and the contribution due to the overlap of two spectral functions at different band indices m and n becomes more pronounced at the Fermi level. Above the minimum, the interband contribution increases till $\eta\sim1$ eV. Because of the finite Wannier orbitals basis, we have the accurate description of energy bands only within the approximate range of $(\epsilon_{F}-10,\epsilon_{F}+5)$ eV for the ferromagnets in consideration. The decreasing trend after $\eta\sim1$ eV is, therefore, an artifact.

\subsection{Moment of inertia}
In Fig. \ref{fig:5}, we report the values for the moment of inertia calculated for bulk Fe, Co and Ni. Analogous to the damping, the inertia tensor is diagonal, resulting in the effective moment of inertia $\mathrm{I} = \mathrm{I}^{xx}+\mathrm{I}^{yy}$.

\begin{figure}[h!]
\includegraphics[width=0.50\textwidth]{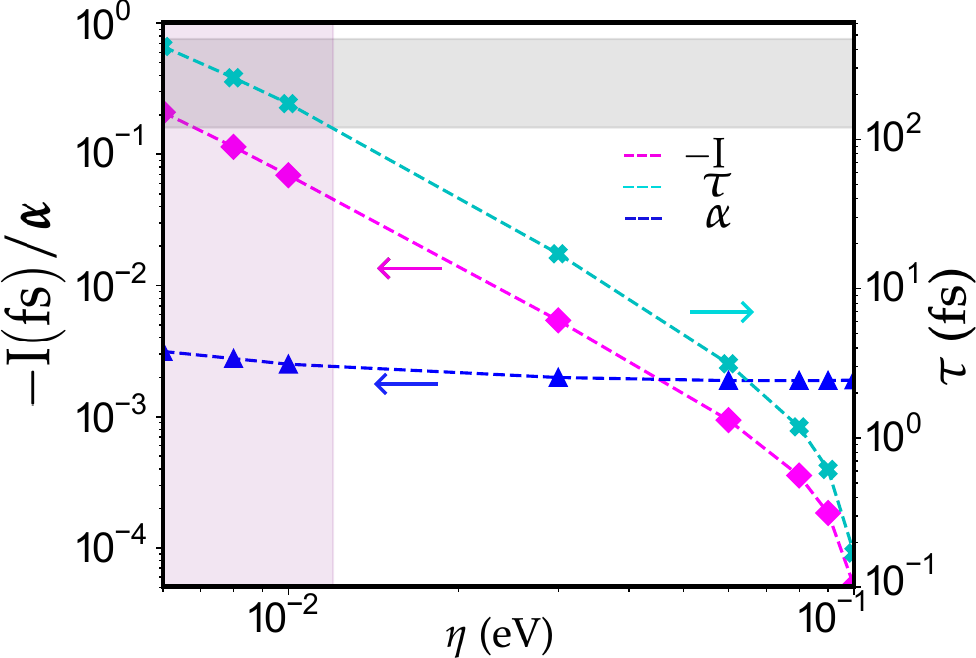}% Here is how to import EPS art
\caption{\label{fig:7} The damping, magnetic moment of inertia and relaxation rate are shown as a function of broadening, $\eta$ in blue, green and red, respectively. The grey-shaded region shows the observed experimental relaxation rate, $\tau$, ranging from $0.12$ to $0.47$ ps. The corresponding range of $\eta$ is shown in purple and is $6-12$ meV. This agrees with the experimental broadening in the range of $5-10$ meV, arising from electron-phonon coupling. The numbers are tabulated in Table \ref{tab:table}}
\end{figure}

The behaviour for $\mathrm{I}$ vs $\eta$ is similar to that of the damping, with smaller and larger $\eta$ trends arising because of intraband and interband contributions, respectively. The overlap term in the moment of inertia is between the $\partial^{2}\mathfrak{R}G/\partial\epsilon^{2}$ and $\mathfrak{I}G$ unlike just $\mathfrak{I}G$ in the damping. In Ref. \cite{thonig2017magnetic}, the moment of inertia is defined in terms of torque matrix elements and the overlap matrix as:
\begin{eqnarray}\label{eqn:28}
{\mathrm{I}}^{\mu \nu }=-\frac{g\hbar }{m_{s}}\sum _{nm}\int {T}_{nm}^{\mu }({\boldsymbol{k}}){T}_{nm}^{\ast \nu }({\boldsymbol{k}}){V}_{nm}{\rm{d}}{\boldsymbol{k}}
\end{eqnarray}
where ${V}_{nm}$ is an overlap function, given by $\int f(\epsilon )({A}_{n{\boldsymbol{k}}}(\epsilon ){B}_{m{\boldsymbol{k}}}(\epsilon )+{B}_{n{\boldsymbol{k}}}(\epsilon ){A}_{m{\boldsymbol{k}}}(\epsilon )){\rm{d}}\epsilon$ and ${B}_{m{\boldsymbol{k}}}(\epsilon )$ is given by $\mathrm{2(}\epsilon -{\epsilon }_{m{\boldsymbol{k}}})({(\epsilon -{\epsilon }_{m{\boldsymbol{k}}})}^{2}-3{{{\eta }}}^{2})/{({(\epsilon -{\epsilon }_{m{\boldsymbol{k}}})}^{2}+{{{\eta }}}^{2})}^{3}$.
 There are other notable features different from the damping. In the limit $\eta\rightarrow0$, the overlap $V_{mn}$ reduces to $2/(\epsilon_{m\bm{k}}-\epsilon_{n\bm{k}})^{3}$ . For intraband transitions ($m=n$), this leads to $\mathrm{I}\rightarrow -\infty$. In the limit $\eta\rightarrow\infty$, $V_{mn}\approx1/\eta^{5}$ which leads to $\mathrm{I}\rightarrow 0$. The behaviour at these two limits is evident from Fig. \ref{fig:5}. The large $\tau$ (small $\eta$) behaviour is consistent with the expression $\mathrm{I} = -\alpha.\tau/2\pi$ derived by F\textrm{$\ddot{a}$}hnle \textit{et al.}\cite{fahnle2013erratum}. Here $\tau$ is the Bloch relaxation lifetime. The behaviour of $\tau$ as a function of $\eta$ using the above expression in the low $\eta$ limit is shown in Fig. \ref{fig:7}. Apart from these limits, the sign change has been observed in a certain range of $\eta$ for Fe and Co. This change in sign can be explained by the Eqn. \eqref{eqn:5}. In the regime of intraband contribution, at a certain $\eta$, the negative and positive terms integrated over $\epsilon$ and $\bm{k}$ become the same, leading to zero inertia. Above that $\eta$, the contribution due to the negative terms decreases until the interband contribution plays a major role leading to maxima in $\mathrm{I}$ (minima in $-\mathrm{I}$). Interband contribution leads to the sign change from + to - and eventually zero at larger $\eta$.

The expression $\mathrm{I} = -\alpha.\tau/2\pi$ derived from the Kambersky model is valid for $\eta<10$ meV, which indicates that damping and moment of inertia have opposite signs. By analyzing the rate of change of magnetic energy, Ref. \cite{bhattacharjee2012atomistic} shows that Gilbert damping and the moment of inertia have opposite signs when magnetization dynamics are sufficiently slow (compared to $\tau$).

 Experimentally, the stiffening of FMR frequency is caused by negative inertia. The softening caused by positive inertia is not observed experimentally. This is because the experimentally realized broadening, $\eta$ caused by electron-phonon scattering and scattering with impurities, is of the order of $5-10$ meV. The values of Bloch relaxation lifetime, $\tau$ measured at the room temperature with the FMR in the high-frequency regime for $\mathrm{Ni_{79}Fe_{21}}$ and Co films of different thickness, range from $0.12-0.47$ ps. The theoretically calculated values for Fe,Ni and Co using the Wannier implementation for the $\eta$ ranging from $5-10$ meV are reported in Table. \ref{tab:table} and lies roughly in the above-mentioned experimental range for the ferromagnetic films.

\section{\label{sec:level5}CONCLUSIONS}

In summary, this paper presents a numerical method to obtain the Gilbert damping and moment of inertia based on the torque-torque correlation model within an \textit{ab initio} Wannier framework. We have also described a technique to calculate the spin-orbit coupling matrix elements via the transformation between the spin-orbit and spin-polarised basis. The damping and inertia calculated using this method for the transition metals like Fe, Co and Ni are in good agreement with the previous studies based on tight binding method\cite{thonig2014gilbert,thonig2017magnetic} and local spin density approximation\cite{PhysRevLett.99.027204}. We have calculated the Bloch relaxation time for the approximate physical range of broadening caused by electron-phonon coupling and lattice defects. The Bloch relaxation time is in good agreement with experimentally reported values using FMR\cite{li2015inertial}. The calculated damping and moment of inertia can be used to study the magnetisation dynamics in the sub-ps regime. In future studies, we plan to use the Wannier implementation to study the contribution of spin pumping terms, arising from the spin currents at the interface of ferromagnetic-normal metal bilayer systems due to the spin-orbit coupling and inversion symmetry breaking to the damping. We also plan to study the magnetic damping and anisotropy in experimentally reported 2D ferromagnetic materials\cite{wang2022two} like $\mathrm{CrGeTe_{3}, CrTe, Cr_{3}Te_{4}}$ etc. The increasing interest in investigating the magnetic properties in 2D ferromagnets is due to magnetic anisotropy, which stabilises the long-range ferromagnetic order in such materials. Moreover, the reduction in dimensionality from bulk to 2D leads to intriguingly distinct magnetic properties compared to the bulk.

\section{\label{sec:level6}ACKNOWLEDGEMENTS}

This work has been supported by a financial grant through the Indo-Korea Science and Technology Center (IKST).
We thank the Supercomputer Education and Research Centre (SERC) at the Indian Institute of Science (IISc) and the Korea Institute
of Science and Technology (KIST) for providing the computational
facilities.

\nocite{*}

\bibliography{main}

\begin{thebibliography}{50}%
\makeatletter
\providecommand \@ifxundefined [1]{%
 \@ifx{#1\undefined}
}%
\providecommand \@ifnum [1]{%
 \ifnum #1\expandafter \@firstoftwo
 \else \expandafter \@secondoftwo
 \fi
}%
\providecommand \@ifx [1]{%
 \ifx #1\expandafter \@firstoftwo
 \else \expandafter \@secondoftwo
 \fi
}%
\providecommand \natexlab [1]{#1}%
\providecommand \enquote  [1]{``#1''}%
\providecommand \bibnamefont  [1]{#1}%
\providecommand \bibfnamefont [1]{#1}%
\providecommand \citenamefont [1]{#1}%
\providecommand \href@noop [0]{\@secondoftwo}%
\providecommand \href [0]{\begingroup \@sanitize@url \@href}%
\providecommand \@href[1]{\@@startlink{#1}\@@href}%
\providecommand \@@href[1]{\endgroup#1\@@endlink}%
\providecommand \@sanitize@url [0]{\catcode `\\12\catcode `\$12\catcode
  `\&12\catcode `\#12\catcode `\^12\catcode `\_12\catcode `\%12\relax}%
\providecommand \@@startlink[1]{}%
\providecommand \@@endlink[0]{}%
\providecommand \url  [0]{\begingroup\@sanitize@url \@url }%
\providecommand \@url [1]{\endgroup\@href {#1}{\urlprefix }}%
\providecommand \urlprefix  [0]{URL }%
\providecommand \Eprint [0]{\href }%
\providecommand \doibase [0]{https://doi.org/}%
\providecommand \selectlanguage [0]{\@gobble}%
\providecommand \bibinfo  [0]{\@secondoftwo}%
\providecommand \bibfield  [0]{\@secondoftwo}%
\providecommand \translation [1]{[#1]}%
\providecommand \BibitemOpen [0]{}%
\providecommand \bibitemStop [0]{}%
\providecommand \bibitemNoStop [0]{.\EOS\space}%
\providecommand \EOS [0]{\spacefactor3000\relax}%
\providecommand \BibitemShut  [1]{\csname bibitem#1\endcsname}%
\let\auto@bib@innerbib\@empty
%</preamble>
\bibitem [{\citenamefont {F{\"a}hnle}\ \emph {et~al.}(2005)\citenamefont
  {F{\"a}hnle}, \citenamefont {Drautz}, \citenamefont {Singer}, \citenamefont
  {Steiauf},\ and\ \citenamefont {Berkov}}]{fahnle2005fast}%
  \BibitemOpen
  \bibfield  {author} {\bibinfo {author} {\bibfnamefont {M.}~\bibnamefont
  {F{\"a}hnle}}, \bibinfo {author} {\bibfnamefont {R.}~\bibnamefont {Drautz}},
  \bibinfo {author} {\bibfnamefont {R.}~\bibnamefont {Singer}}, \bibinfo
  {author} {\bibfnamefont {D.}~\bibnamefont {Steiauf}},\ and\ \bibinfo {author}
  {\bibfnamefont {D.}~\bibnamefont {Berkov}},\ }\bibfield  {title} {\bibinfo
  {title} {A fast ab initio approach to the simulation of spin dynamics},\
  }\href@noop {} {\bibfield  {journal} {\bibinfo  {journal} {Computational
  Materials Science}\ }\textbf {\bibinfo {volume} {32}},\ \bibinfo {pages}
  {118} (\bibinfo {year} {2005})}\BibitemShut {NoStop}%
\bibitem [{\citenamefont {Antropov}\ \emph {et~al.}(1995)\citenamefont
  {Antropov}, \citenamefont {Katsnelson}, \citenamefont {Van~Schilfgaarde},\
  and\ \citenamefont {Harmon}}]{antropov1995ab}%
  \BibitemOpen
  \bibfield  {author} {\bibinfo {author} {\bibfnamefont {V.~P.}\ \bibnamefont
  {Antropov}}, \bibinfo {author} {\bibfnamefont {M.}~\bibnamefont
  {Katsnelson}}, \bibinfo {author} {\bibfnamefont {M.}~\bibnamefont
  {Van~Schilfgaarde}},\ and\ \bibinfo {author} {\bibfnamefont {B.}~\bibnamefont
  {Harmon}},\ }\bibfield  {title} {\bibinfo {title} {Ab initio spin dynamics in
  magnets},\ }\href@noop {} {\bibfield  {journal} {\bibinfo  {journal}
  {Physical Review Letters}\ }\textbf {\bibinfo {volume} {75}},\ \bibinfo
  {pages} {729} (\bibinfo {year} {1995})}\BibitemShut {NoStop}%
\bibitem [{\citenamefont {Skubic}\ \emph {et~al.}(2008)\citenamefont {Skubic},
  \citenamefont {Hellsvik}, \citenamefont {Nordstr{\"o}m},\ and\ \citenamefont
  {Eriksson}}]{skubic2008method}%
  \BibitemOpen
  \bibfield  {author} {\bibinfo {author} {\bibfnamefont {B.}~\bibnamefont
  {Skubic}}, \bibinfo {author} {\bibfnamefont {J.}~\bibnamefont {Hellsvik}},
  \bibinfo {author} {\bibfnamefont {L.}~\bibnamefont {Nordstr{\"o}m}},\ and\
  \bibinfo {author} {\bibfnamefont {O.}~\bibnamefont {Eriksson}},\ }\bibfield
  {title} {\bibinfo {title} {A method for atomistic spin dynamics simulations:
  implementation and examples},\ }\href@noop {} {\bibfield  {journal} {\bibinfo
   {journal} {Journal of Physics: Condensed Matter}\ }\textbf {\bibinfo
  {volume} {20}},\ \bibinfo {pages} {315203} (\bibinfo {year}
  {2008})}\BibitemShut {NoStop}%
\bibitem [{\citenamefont {Antropov}\ \emph {et~al.}(1996)\citenamefont
  {Antropov}, \citenamefont {Katsnelson}, \citenamefont {Harmon}, \citenamefont
  {Van~Schilfgaarde},\ and\ \citenamefont {Kusnezov}}]{antropov1996spin}%
  \BibitemOpen
  \bibfield  {author} {\bibinfo {author} {\bibfnamefont {V.}~\bibnamefont
  {Antropov}}, \bibinfo {author} {\bibfnamefont {M.}~\bibnamefont
  {Katsnelson}}, \bibinfo {author} {\bibfnamefont {B.}~\bibnamefont {Harmon}},
  \bibinfo {author} {\bibfnamefont {M.}~\bibnamefont {Van~Schilfgaarde}},\ and\
  \bibinfo {author} {\bibfnamefont {D.}~\bibnamefont {Kusnezov}},\ }\bibfield
  {title} {\bibinfo {title} {Spin dynamics in magnets: Equation of motion and
  finite temperature effects},\ }\href@noop {} {\bibfield  {journal} {\bibinfo
  {journal} {Physical Review B}\ }\textbf {\bibinfo {volume} {54}},\ \bibinfo
  {pages} {1019} (\bibinfo {year} {1996})}\BibitemShut {NoStop}%
\bibitem [{\citenamefont {Steiauf}\ and\ \citenamefont
  {F{\"a}hnle}(2005)}]{steiauf2005damping}%
  \BibitemOpen
  \bibfield  {author} {\bibinfo {author} {\bibfnamefont {D.}~\bibnamefont
  {Steiauf}}\ and\ \bibinfo {author} {\bibfnamefont {M.}~\bibnamefont
  {F{\"a}hnle}},\ }\bibfield  {title} {\bibinfo {title} {Damping of spin
  dynamics in nanostructures: An ab initio study},\ }\href@noop {} {\bibfield
  {journal} {\bibinfo  {journal} {Physical Review B}\ }\textbf {\bibinfo
  {volume} {72}},\ \bibinfo {pages} {064450} (\bibinfo {year}
  {2005})}\BibitemShut {NoStop}%
\bibitem [{\citenamefont {Parkin}\ \emph {et~al.}(2003)\citenamefont {Parkin},
  \citenamefont {Jiang}, \citenamefont {Kaiser}, \citenamefont {Panchula},
  \citenamefont {Roche},\ and\ \citenamefont
  {Samant}}]{parkin2003magnetically}%
  \BibitemOpen
  \bibfield  {author} {\bibinfo {author} {\bibfnamefont {S.}~\bibnamefont
  {Parkin}}, \bibinfo {author} {\bibfnamefont {X.}~\bibnamefont {Jiang}},
  \bibinfo {author} {\bibfnamefont {C.}~\bibnamefont {Kaiser}}, \bibinfo
  {author} {\bibfnamefont {A.}~\bibnamefont {Panchula}}, \bibinfo {author}
  {\bibfnamefont {K.}~\bibnamefont {Roche}},\ and\ \bibinfo {author}
  {\bibfnamefont {M.}~\bibnamefont {Samant}},\ }\bibfield  {title} {\bibinfo
  {title} {Magnetically engineered spintronic sensors and memory},\ }\href@noop
  {} {\bibfield  {journal} {\bibinfo  {journal} {Proceedings of the IEEE}\
  }\textbf {\bibinfo {volume} {91}},\ \bibinfo {pages} {661} (\bibinfo {year}
  {2003})}\BibitemShut {NoStop}%
\bibitem [{\citenamefont {Xu}\ and\ \citenamefont
  {Thompson}(2006)}]{xu2006spintronic}%
  \BibitemOpen
  \bibfield  {author} {\bibinfo {author} {\bibfnamefont {Y.}~\bibnamefont
  {Xu}}\ and\ \bibinfo {author} {\bibfnamefont {S.}~\bibnamefont {Thompson}},\
  }\href@noop {} {\emph {\bibinfo {title} {Spintronic materials and
  technology}}}\ (\bibinfo  {publisher} {CRC press},\ \bibinfo {year}
  {2006})\BibitemShut {NoStop}%
\bibitem [{\citenamefont {Kim}\ and\ \citenamefont
  {Lee}(2016)}]{kim2016chiral}%
  \BibitemOpen
  \bibfield  {author} {\bibinfo {author} {\bibfnamefont {K.-W.}\ \bibnamefont
  {Kim}}\ and\ \bibinfo {author} {\bibfnamefont {H.-W.}\ \bibnamefont {Lee}},\
  }\bibfield  {title} {\bibinfo {title} {Chiral damping},\ }\href@noop {}
  {\bibfield  {journal} {\bibinfo  {journal} {Nature Materials}\ }\textbf
  {\bibinfo {volume} {15}},\ \bibinfo {pages} {253} (\bibinfo {year}
  {2016})}\BibitemShut {NoStop}%
\bibitem [{\citenamefont {Chumak}\ \emph {et~al.}(2015)\citenamefont {Chumak},
  \citenamefont {Vasyuchka}, \citenamefont {Serga},\ and\ \citenamefont
  {Hillebrands}}]{chumak2015magnon}%
  \BibitemOpen
  \bibfield  {author} {\bibinfo {author} {\bibfnamefont {A.~V.}\ \bibnamefont
  {Chumak}}, \bibinfo {author} {\bibfnamefont {V.~I.}\ \bibnamefont
  {Vasyuchka}}, \bibinfo {author} {\bibfnamefont {A.~A.}\ \bibnamefont
  {Serga}},\ and\ \bibinfo {author} {\bibfnamefont {B.}~\bibnamefont
  {Hillebrands}},\ }\bibfield  {title} {\bibinfo {title} {Magnon spintronics},\
  }\href@noop {} {\bibfield  {journal} {\bibinfo  {journal} {Nature Physics}\
  }\textbf {\bibinfo {volume} {11}},\ \bibinfo {pages} {453} (\bibinfo {year}
  {2015})}\BibitemShut {NoStop}%
\bibitem [{\citenamefont {Gilbert}(2004)}]{gilbert2004phenomenological}%
  \BibitemOpen
  \bibfield  {author} {\bibinfo {author} {\bibfnamefont {T.~L.}\ \bibnamefont
  {Gilbert}},\ }\bibfield  {title} {\bibinfo {title} {A phenomenological theory
  of damping in ferromagnetic materials},\ }\href@noop {} {\bibfield  {journal}
  {\bibinfo  {journal} {IEEE Transactions on Magnetics}\ }\textbf {\bibinfo
  {volume} {40}},\ \bibinfo {pages} {3443} (\bibinfo {year}
  {2004})}\BibitemShut {NoStop}%
\bibitem [{\citenamefont {Bhattacharjee}\ \emph {et~al.}(2012)\citenamefont
  {Bhattacharjee}, \citenamefont {Nordstr{\"o}m},\ and\ \citenamefont
  {Fransson}}]{bhattacharjee2012atomistic}%
  \BibitemOpen
  \bibfield  {author} {\bibinfo {author} {\bibfnamefont {S.}~\bibnamefont
  {Bhattacharjee}}, \bibinfo {author} {\bibfnamefont {L.}~\bibnamefont
  {Nordstr{\"o}m}},\ and\ \bibinfo {author} {\bibfnamefont {J.}~\bibnamefont
  {Fransson}},\ }\bibfield  {title} {\bibinfo {title} {Atomistic spin dynamic
  method with both damping and moment of inertia effects included from first
  principles},\ }\href@noop {} {\bibfield  {journal} {\bibinfo  {journal}
  {Physical Review Letters}\ }\textbf {\bibinfo {volume} {108}},\ \bibinfo
  {pages} {057204} (\bibinfo {year} {2012})}\BibitemShut {NoStop}%
\bibitem [{\citenamefont {Ciornei}\ \emph {et~al.}(2011)\citenamefont
  {Ciornei}, \citenamefont {Rub{\'\i}},\ and\ \citenamefont
  {Wegrowe}}]{ciornei2011magnetization}%
  \BibitemOpen
  \bibfield  {author} {\bibinfo {author} {\bibfnamefont {M.-C.}\ \bibnamefont
  {Ciornei}}, \bibinfo {author} {\bibfnamefont {J.}~\bibnamefont {Rub{\'\i}}},\
  and\ \bibinfo {author} {\bibfnamefont {J.-E.}\ \bibnamefont {Wegrowe}},\
  }\bibfield  {title} {\bibinfo {title} {Magnetization dynamics in the inertial
  regime: Nutation predicted at short time scales},\ }\href@noop {} {\bibfield
  {journal} {\bibinfo  {journal} {Physical Review B}\ }\textbf {\bibinfo
  {volume} {83}},\ \bibinfo {pages} {020410} (\bibinfo {year}
  {2011})}\BibitemShut {NoStop}%
\bibitem [{\citenamefont {B{\"o}ttcher}\ and\ \citenamefont
  {Henk}(2012)}]{bottcher2012significance}%
  \BibitemOpen
  \bibfield  {author} {\bibinfo {author} {\bibfnamefont {D.}~\bibnamefont
  {B{\"o}ttcher}}\ and\ \bibinfo {author} {\bibfnamefont {J.}~\bibnamefont
  {Henk}},\ }\bibfield  {title} {\bibinfo {title} {Significance of nutation in
  magnetization dynamics of nanostructures},\ }\href@noop {} {\bibfield
  {journal} {\bibinfo  {journal} {Physical Review B}\ }\textbf {\bibinfo
  {volume} {86}},\ \bibinfo {pages} {020404} (\bibinfo {year}
  {2012})}\BibitemShut {NoStop}%
\bibitem [{\citenamefont {Wieser}(2013)}]{wieser2013comparison}%
  \BibitemOpen
  \bibfield  {author} {\bibinfo {author} {\bibfnamefont {R.}~\bibnamefont
  {Wieser}},\ }\bibfield  {title} {\bibinfo {title} {Comparison of quantum and
  classical relaxation in spin dynamics},\ }\href@noop {} {\bibfield  {journal}
  {\bibinfo  {journal} {Physical Review Letters}\ }\textbf {\bibinfo {volume}
  {110}},\ \bibinfo {pages} {147201} (\bibinfo {year} {2013})}\BibitemShut
  {NoStop}%
\bibitem [{\citenamefont {Chudnovskiy}\ \emph {et~al.}(2014)\citenamefont
  {Chudnovskiy}, \citenamefont {H{\"u}bner}, \citenamefont {Baxevanis},\ and\
  \citenamefont {Pfannkuche}}]{chudnovskiy2014spin}%
  \BibitemOpen
  \bibfield  {author} {\bibinfo {author} {\bibfnamefont {A.}~\bibnamefont
  {Chudnovskiy}}, \bibinfo {author} {\bibfnamefont {C.}~\bibnamefont
  {H{\"u}bner}}, \bibinfo {author} {\bibfnamefont {B.}~\bibnamefont
  {Baxevanis}},\ and\ \bibinfo {author} {\bibfnamefont {D.}~\bibnamefont
  {Pfannkuche}},\ }\bibfield  {title} {\bibinfo {title} {Spin switching: From
  quantum to quasiclassical approach},\ }\href@noop {} {\bibfield  {journal}
  {\bibinfo  {journal} {physica status solidi (b)}\ }\textbf {\bibinfo {volume}
  {251}},\ \bibinfo {pages} {1764} (\bibinfo {year} {2014})}\BibitemShut
  {NoStop}%
\bibitem [{\citenamefont {Fuchs}\ \emph {et~al.}(2007)\citenamefont {Fuchs},
  \citenamefont {Sankey}, \citenamefont {Pribiag}, \citenamefont {Qian},
  \citenamefont {Braganca}, \citenamefont {Garcia}, \citenamefont {Ryan},
  \citenamefont {Li}, \citenamefont {Ozatay}, \citenamefont {Ralph} \emph
  {et~al.}}]{fuchs2007spin}%
  \BibitemOpen
  \bibfield  {author} {\bibinfo {author} {\bibfnamefont {G.}~\bibnamefont
  {Fuchs}}, \bibinfo {author} {\bibfnamefont {J.}~\bibnamefont {Sankey}},
  \bibinfo {author} {\bibfnamefont {V.}~\bibnamefont {Pribiag}}, \bibinfo
  {author} {\bibfnamefont {L.}~\bibnamefont {Qian}}, \bibinfo {author}
  {\bibfnamefont {P.}~\bibnamefont {Braganca}}, \bibinfo {author}
  {\bibfnamefont {A.}~\bibnamefont {Garcia}}, \bibinfo {author} {\bibfnamefont
  {E.}~\bibnamefont {Ryan}}, \bibinfo {author} {\bibfnamefont {Z.-P.}\
  \bibnamefont {Li}}, \bibinfo {author} {\bibfnamefont {O.}~\bibnamefont
  {Ozatay}}, \bibinfo {author} {\bibfnamefont {D.}~\bibnamefont {Ralph}}, \emph
  {et~al.},\ }\bibfield  {title} {\bibinfo {title} {Spin-torque ferromagnetic
  resonance measurements of damping in nanomagnets},\ }\href@noop {} {\bibfield
   {journal} {\bibinfo  {journal} {Applied Physics Letters}\ }\textbf {\bibinfo
  {volume} {91}} (\bibinfo {year} {2007})}\BibitemShut {NoStop}%
\bibitem [{\citenamefont {Oogane}\ \emph {et~al.}(2006)\citenamefont {Oogane},
  \citenamefont {Wakitani}, \citenamefont {Yakata}, \citenamefont {Yilgin},
  \citenamefont {Ando}, \citenamefont {Sakuma},\ and\ \citenamefont
  {Miyazaki}}]{oogane2006magnetic}%
  \BibitemOpen
  \bibfield  {author} {\bibinfo {author} {\bibfnamefont {M.}~\bibnamefont
  {Oogane}}, \bibinfo {author} {\bibfnamefont {T.}~\bibnamefont {Wakitani}},
  \bibinfo {author} {\bibfnamefont {S.}~\bibnamefont {Yakata}}, \bibinfo
  {author} {\bibfnamefont {R.}~\bibnamefont {Yilgin}}, \bibinfo {author}
  {\bibfnamefont {Y.}~\bibnamefont {Ando}}, \bibinfo {author} {\bibfnamefont
  {A.}~\bibnamefont {Sakuma}},\ and\ \bibinfo {author} {\bibfnamefont
  {T.}~\bibnamefont {Miyazaki}},\ }\bibfield  {title} {\bibinfo {title}
  {Magnetic damping in ferromagnetic thin films},\ }\href@noop {} {\bibfield
  {journal} {\bibinfo  {journal} {Japanese Journal of Applied Physics}\
  }\textbf {\bibinfo {volume} {45}},\ \bibinfo {pages} {3889} (\bibinfo {year}
  {2006})}\BibitemShut {NoStop}%
\bibitem [{\citenamefont {Barati}\ \emph {et~al.}(2013)\citenamefont {Barati},
  \citenamefont {Cinal}, \citenamefont {Edwards},\ and\ \citenamefont
  {Umerski}}]{barati2013calculation}%
  \BibitemOpen
  \bibfield  {author} {\bibinfo {author} {\bibfnamefont {E.}~\bibnamefont
  {Barati}}, \bibinfo {author} {\bibfnamefont {M.}~\bibnamefont {Cinal}},
  \bibinfo {author} {\bibfnamefont {D.}~\bibnamefont {Edwards}},\ and\ \bibinfo
  {author} {\bibfnamefont {A.}~\bibnamefont {Umerski}},\ }\bibfield  {title}
  {\bibinfo {title} {Calculation of {G}ilbert damping in ferromagnetic films},\
  }in\ \href@noop {} {\emph {\bibinfo {booktitle} {EPJ Web of Conferences}}},\
  Vol.~\bibinfo {volume} {40}\ (\bibinfo {organization} {EDP Sciences},\
  \bibinfo {year} {2013})\ p.\ \bibinfo {pages} {18003}\BibitemShut {NoStop}%
\bibitem [{\citenamefont {Bhagat}\ and\ \citenamefont
  {Lubitz}(1974)}]{bhagat1974temperature}%
  \BibitemOpen
  \bibfield  {author} {\bibinfo {author} {\bibfnamefont {S.}~\bibnamefont
  {Bhagat}}\ and\ \bibinfo {author} {\bibfnamefont {P.}~\bibnamefont
  {Lubitz}},\ }\bibfield  {title} {\bibinfo {title} {Temperature variation of
  ferromagnetic relaxation in the 3d transition metals},\ }\href@noop {}
  {\bibfield  {journal} {\bibinfo  {journal} {Physical Review B}\ }\textbf
  {\bibinfo {volume} {10}},\ \bibinfo {pages} {179} (\bibinfo {year}
  {1974})}\BibitemShut {NoStop}%
\bibitem [{\citenamefont {Schreiber}\ \emph {et~al.}(1995)\citenamefont
  {Schreiber}, \citenamefont {Pflaum}, \citenamefont {Frait}, \citenamefont
  {M{\"u}hge},\ and\ \citenamefont {Pelzl}}]{schreiber1995gilbert}%
  \BibitemOpen
  \bibfield  {author} {\bibinfo {author} {\bibfnamefont {F.}~\bibnamefont
  {Schreiber}}, \bibinfo {author} {\bibfnamefont {J.}~\bibnamefont {Pflaum}},
  \bibinfo {author} {\bibfnamefont {Z.}~\bibnamefont {Frait}}, \bibinfo
  {author} {\bibfnamefont {T.}~\bibnamefont {M{\"u}hge}},\ and\ \bibinfo
  {author} {\bibfnamefont {J.}~\bibnamefont {Pelzl}},\ }\bibfield  {title}
  {\bibinfo {title} {Gilbert damping and g-factor in {F}ex{C}o1-x alloy
  films},\ }\href@noop {} {\bibfield  {journal} {\bibinfo  {journal} {Solid
  State Communications}\ }\textbf {\bibinfo {volume} {93}},\ \bibinfo {pages}
  {965} (\bibinfo {year} {1995})}\BibitemShut {NoStop}%
\bibitem [{\citenamefont {Inaba}\ \emph {et~al.}(2006)\citenamefont {Inaba},
  \citenamefont {Asanuma}, \citenamefont {Igarashi}, \citenamefont {Mori},
  \citenamefont {Kirino}, \citenamefont {Koike},\ and\ \citenamefont
  {Morita}}]{inaba2006damping}%
  \BibitemOpen
  \bibfield  {author} {\bibinfo {author} {\bibfnamefont {N.}~\bibnamefont
  {Inaba}}, \bibinfo {author} {\bibfnamefont {H.}~\bibnamefont {Asanuma}},
  \bibinfo {author} {\bibfnamefont {S.}~\bibnamefont {Igarashi}}, \bibinfo
  {author} {\bibfnamefont {S.}~\bibnamefont {Mori}}, \bibinfo {author}
  {\bibfnamefont {F.}~\bibnamefont {Kirino}}, \bibinfo {author} {\bibfnamefont
  {K.}~\bibnamefont {Koike}},\ and\ \bibinfo {author} {\bibfnamefont
  {H.}~\bibnamefont {Morita}},\ }\bibfield  {title} {\bibinfo {title} {Damping
  constants of {N}i-{F}e and {N}i-{C}o alloy thin films},\ }\href@noop {}
  {\bibfield  {journal} {\bibinfo  {journal} {IEEE Transactions on Magnetics}\
  }\textbf {\bibinfo {volume} {42}},\ \bibinfo {pages} {2372} (\bibinfo {year}
  {2006})}\BibitemShut {NoStop}%
\bibitem [{\citenamefont {Song}\ \emph {et~al.}(2013)\citenamefont {Song},
  \citenamefont {Lee}, \citenamefont {Sohn}, \citenamefont {Yang},
  \citenamefont {Parkin}, \citenamefont {You},\ and\ \citenamefont
  {Shin}}]{song2013observation}%
  \BibitemOpen
  \bibfield  {author} {\bibinfo {author} {\bibfnamefont {H.-S.}\ \bibnamefont
  {Song}}, \bibinfo {author} {\bibfnamefont {K.-D.}\ \bibnamefont {Lee}},
  \bibinfo {author} {\bibfnamefont {J.-W.}\ \bibnamefont {Sohn}}, \bibinfo
  {author} {\bibfnamefont {S.-H.}\ \bibnamefont {Yang}}, \bibinfo {author}
  {\bibfnamefont {S.~S.}\ \bibnamefont {Parkin}}, \bibinfo {author}
  {\bibfnamefont {C.-Y.}\ \bibnamefont {You}},\ and\ \bibinfo {author}
  {\bibfnamefont {S.-C.}\ \bibnamefont {Shin}},\ }\bibfield  {title} {\bibinfo
  {title} {Observation of the intrinsic {G}ilbert damping constant in {C}o/{N}i
  multilayers independent of the stack number with perpendicular anisotropy},\
  }\href@noop {} {\bibfield  {journal} {\bibinfo  {journal} {Applied Physics
  Letters}\ }\textbf {\bibinfo {volume} {102}} (\bibinfo {year}
  {2013})}\BibitemShut {NoStop}%
\bibitem [{\citenamefont {Mizukami}\ \emph {et~al.}(2010)\citenamefont
  {Mizukami}, \citenamefont {Sajitha}, \citenamefont {Watanabe}, \citenamefont
  {Wu}, \citenamefont {Miyazaki}, \citenamefont {Naganuma}, \citenamefont
  {Oogane},\ and\ \citenamefont {Ando}}]{mizukami2010gilbert}%
  \BibitemOpen
  \bibfield  {author} {\bibinfo {author} {\bibfnamefont {S.}~\bibnamefont
  {Mizukami}}, \bibinfo {author} {\bibfnamefont {E.}~\bibnamefont {Sajitha}},
  \bibinfo {author} {\bibfnamefont {D.}~\bibnamefont {Watanabe}}, \bibinfo
  {author} {\bibfnamefont {F.}~\bibnamefont {Wu}}, \bibinfo {author}
  {\bibfnamefont {T.}~\bibnamefont {Miyazaki}}, \bibinfo {author}
  {\bibfnamefont {H.}~\bibnamefont {Naganuma}}, \bibinfo {author}
  {\bibfnamefont {M.}~\bibnamefont {Oogane}},\ and\ \bibinfo {author}
  {\bibfnamefont {Y.}~\bibnamefont {Ando}},\ }\bibfield  {title} {\bibinfo
  {title} {Gilbert damping in perpendicularly magnetized {P}t/{C}o/{P}t films
  investigated by all-optical pump-probe technique},\ }\href@noop {} {\bibfield
   {journal} {\bibinfo  {journal} {Applied Physics Letters}\ }\textbf {\bibinfo
  {volume} {96}} (\bibinfo {year} {2010})}\BibitemShut {NoStop}%
\bibitem [{\citenamefont {Trunova}(2009)}]{trunova2009ferromagnetische}%
  \BibitemOpen
  \bibfield  {author} {\bibinfo {author} {\bibfnamefont {A.}~\bibnamefont
  {Trunova}},\ }\emph {\bibinfo {title} {Ferromagnetische resonanz an
  oxidfreien magnetischen {F}e und {F}e{R}h nanopartikeln}},\ \href@noop {}
  {Ph.D. thesis} (\bibinfo {year} {2009})\BibitemShut {NoStop}%
\bibitem [{\citenamefont {Heinrich}\ and\ \citenamefont
  {Frait}(1966)}]{heinrich1966temperature}%
  \BibitemOpen
  \bibfield  {author} {\bibinfo {author} {\bibfnamefont {B.}~\bibnamefont
  {Heinrich}}\ and\ \bibinfo {author} {\bibfnamefont {Z.}~\bibnamefont
  {Frait}},\ }\bibfield  {title} {\bibinfo {title} {Temperature {D}ependence of
  the {FMR} {L}inewidth of {I}ron {S}ingle-{C}rystal {P}latelets},\ }\href@noop
  {} {\bibfield  {journal} {\bibinfo  {journal} {Physica Status Solidi (b)}\
  }\textbf {\bibinfo {volume} {16}},\ \bibinfo {pages} {K11} (\bibinfo {year}
  {1966})}\BibitemShut {NoStop}%
\bibitem [{\citenamefont {Zhao}\ \emph {et~al.}(2016)\citenamefont {Zhao},
  \citenamefont {Song}, \citenamefont {Yang}, \citenamefont {Su}, \citenamefont
  {Yuan}, \citenamefont {Parkin}, \citenamefont {Shi},\ and\ \citenamefont
  {Han}}]{zhao2016experimental}%
  \BibitemOpen
  \bibfield  {author} {\bibinfo {author} {\bibfnamefont {Y.}~\bibnamefont
  {Zhao}}, \bibinfo {author} {\bibfnamefont {Q.}~\bibnamefont {Song}}, \bibinfo
  {author} {\bibfnamefont {S.-H.}\ \bibnamefont {Yang}}, \bibinfo {author}
  {\bibfnamefont {T.}~\bibnamefont {Su}}, \bibinfo {author} {\bibfnamefont
  {W.}~\bibnamefont {Yuan}}, \bibinfo {author} {\bibfnamefont {S.~S.}\
  \bibnamefont {Parkin}}, \bibinfo {author} {\bibfnamefont {J.}~\bibnamefont
  {Shi}},\ and\ \bibinfo {author} {\bibfnamefont {W.}~\bibnamefont {Han}},\
  }\bibfield  {title} {\bibinfo {title} {Experimental investigation of
  temperature-dependent {G}ilbert damping in permalloy thin films},\
  }\href@noop {} {\bibfield  {journal} {\bibinfo  {journal} {Scientific
  Reports}\ }\textbf {\bibinfo {volume} {6}},\ \bibinfo {pages} {1} (\bibinfo
  {year} {2016})}\BibitemShut {NoStop}%
\bibitem [{\citenamefont {Li}\ \emph {et~al.}(2015)\citenamefont {Li},
  \citenamefont {Barra}, \citenamefont {Auffret}, \citenamefont {Ebels},\ and\
  \citenamefont {Bailey}}]{li2015inertial}%
  \BibitemOpen
  \bibfield  {author} {\bibinfo {author} {\bibfnamefont {Y.}~\bibnamefont
  {Li}}, \bibinfo {author} {\bibfnamefont {A.-L.}\ \bibnamefont {Barra}},
  \bibinfo {author} {\bibfnamefont {S.}~\bibnamefont {Auffret}}, \bibinfo
  {author} {\bibfnamefont {U.}~\bibnamefont {Ebels}},\ and\ \bibinfo {author}
  {\bibfnamefont {W.~E.}\ \bibnamefont {Bailey}},\ }\bibfield  {title}
  {\bibinfo {title} {Inertial terms to magnetization dynamics in ferromagnetic
  thin films},\ }\href@noop {} {\bibfield  {journal} {\bibinfo  {journal}
  {Physical Review B}\ }\textbf {\bibinfo {volume} {92}},\ \bibinfo {pages}
  {140413} (\bibinfo {year} {2015})}\BibitemShut {NoStop}%
\bibitem [{\citenamefont {Umetsu}\ \emph {et~al.}(2012)\citenamefont {Umetsu},
  \citenamefont {Miura},\ and\ \citenamefont {Sakuma}}]{umetsu2012theoretical}%
  \BibitemOpen
  \bibfield  {author} {\bibinfo {author} {\bibfnamefont {N.}~\bibnamefont
  {Umetsu}}, \bibinfo {author} {\bibfnamefont {D.}~\bibnamefont {Miura}},\ and\
  \bibinfo {author} {\bibfnamefont {A.}~\bibnamefont {Sakuma}},\ }\bibfield
  {title} {\bibinfo {title} {Theoretical study on {G}ilbert damping of
  nonuniform magnetization precession in ferromagnetic metals},\ }\href@noop {}
  {\bibfield  {journal} {\bibinfo  {journal} {Journal of the Physical Society
  of Japan}\ }\textbf {\bibinfo {volume} {81}},\ \bibinfo {pages} {114716}
  (\bibinfo {year} {2012})}\BibitemShut {NoStop}%
\bibitem [{\citenamefont {Thonig}\ and\ \citenamefont
  {Henk}(2014)}]{thonig2014gilbert}%
  \BibitemOpen
  \bibfield  {author} {\bibinfo {author} {\bibfnamefont {D.}~\bibnamefont
  {Thonig}}\ and\ \bibinfo {author} {\bibfnamefont {J.}~\bibnamefont {Henk}},\
  }\bibfield  {title} {\bibinfo {title} {Gilbert damping tensor within the
  breathing fermi surface model: anisotropy and non-locality},\ }\href@noop {}
  {\bibfield  {journal} {\bibinfo  {journal} {New Journal of Physics}\ }\textbf
  {\bibinfo {volume} {16}},\ \bibinfo {pages} {013032} (\bibinfo {year}
  {2014})}\BibitemShut {NoStop}%
\bibitem [{\citenamefont {Thonig}\ \emph {et~al.}(2017)\citenamefont {Thonig},
  \citenamefont {Eriksson},\ and\ \citenamefont
  {Pereiro}}]{thonig2017magnetic}%
  \BibitemOpen
  \bibfield  {author} {\bibinfo {author} {\bibfnamefont {D.}~\bibnamefont
  {Thonig}}, \bibinfo {author} {\bibfnamefont {O.}~\bibnamefont {Eriksson}},\
  and\ \bibinfo {author} {\bibfnamefont {M.}~\bibnamefont {Pereiro}},\
  }\bibfield  {title} {\bibinfo {title} {Magnetic moment of inertia within the
  torque-torque correlation model},\ }\href@noop {} {\bibfield  {journal}
  {\bibinfo  {journal} {Scientific Reports}\ }\textbf {\bibinfo {volume} {7}},\
  \bibinfo {pages} {931} (\bibinfo {year} {2017})}\BibitemShut {NoStop}%
\bibitem [{\citenamefont {Thonig}\ \emph {et~al.}(2018)\citenamefont {Thonig},
  \citenamefont {Kvashnin}, \citenamefont {Eriksson},\ and\ \citenamefont
  {Pereiro}}]{thonig2018nonlocal}%
  \BibitemOpen
  \bibfield  {author} {\bibinfo {author} {\bibfnamefont {D.}~\bibnamefont
  {Thonig}}, \bibinfo {author} {\bibfnamefont {Y.}~\bibnamefont {Kvashnin}},
  \bibinfo {author} {\bibfnamefont {O.}~\bibnamefont {Eriksson}},\ and\
  \bibinfo {author} {\bibfnamefont {M.}~\bibnamefont {Pereiro}},\ }\bibfield
  {title} {\bibinfo {title} {Nonlocal {G}ilbert damping tensor within the
  torque-torque correlation model},\ }\href@noop {} {\bibfield  {journal}
  {\bibinfo  {journal} {Physical Review Materials}\ }\textbf {\bibinfo {volume}
  {2}},\ \bibinfo {pages} {013801} (\bibinfo {year} {2018})}\BibitemShut
  {NoStop}%
\bibitem [{\citenamefont {Gilmore}\ \emph {et~al.}(2007)\citenamefont
  {Gilmore}, \citenamefont {Idzerda},\ and\ \citenamefont
  {Stiles}}]{PhysRevLett.99.027204}%
  \BibitemOpen
  \bibfield  {author} {\bibinfo {author} {\bibfnamefont {K.}~\bibnamefont
  {Gilmore}}, \bibinfo {author} {\bibfnamefont {Y.~U.}\ \bibnamefont
  {Idzerda}},\ and\ \bibinfo {author} {\bibfnamefont {M.~D.}\ \bibnamefont
  {Stiles}},\ }\bibfield  {title} {\bibinfo {title} {Identification of the
  {D}ominant {P}recession-{D}amping {M}echanism in {F}e, {C}o, and {N}i by
  {F}irst-{P}rinciples {C}alculations},\ }\href
  {https://doi.org/10.1103/PhysRevLett.99.027204} {\bibfield  {journal}
  {\bibinfo  {journal} {Phys. Rev. Lett.}\ }\textbf {\bibinfo {volume} {99}},\
  \bibinfo {pages} {027204} (\bibinfo {year} {2007})}\BibitemShut {NoStop}%
\bibitem [{\citenamefont {Ebert}\ \emph {et~al.}(2011)\citenamefont {Ebert},
  \citenamefont {Mankovsky}, \citenamefont {K\"odderitzsch},\ and\
  \citenamefont {Kelly}}]{PhysRevLett.107.066603}%
  \BibitemOpen
  \bibfield  {author} {\bibinfo {author} {\bibfnamefont {H.}~\bibnamefont
  {Ebert}}, \bibinfo {author} {\bibfnamefont {S.}~\bibnamefont {Mankovsky}},
  \bibinfo {author} {\bibfnamefont {D.}~\bibnamefont {K\"odderitzsch}},\ and\
  \bibinfo {author} {\bibfnamefont {P.~J.}\ \bibnamefont {Kelly}},\ }\bibfield
  {title} {\bibinfo {title} {Ab {I}nitio {C}alculation of the {G}ilbert
  {D}amping {P}arameter via the {L}inear {R}esponse {F}ormalism},\ }\href
  {https://doi.org/10.1103/PhysRevLett.107.066603} {\bibfield  {journal}
  {\bibinfo  {journal} {Phys. Rev. Lett.}\ }\textbf {\bibinfo {volume} {107}},\
  \bibinfo {pages} {066603} (\bibinfo {year} {2011})}\BibitemShut {NoStop}%
\bibitem [{\citenamefont {Gilmore}\ and\ \citenamefont
  {Stiles}(2009)}]{gilmore2009evaluating}%
  \BibitemOpen
  \bibfield  {author} {\bibinfo {author} {\bibfnamefont {K.}~\bibnamefont
  {Gilmore}}\ and\ \bibinfo {author} {\bibfnamefont {M.~D.}\ \bibnamefont
  {Stiles}},\ }\bibfield  {title} {\bibinfo {title} {Evaluating the locality of
  intrinsic precession damping in transition metals},\ }\href@noop {}
  {\bibfield  {journal} {\bibinfo  {journal} {Physical Review B}\ }\textbf
  {\bibinfo {volume} {79}},\ \bibinfo {pages} {132407} (\bibinfo {year}
  {2009})}\BibitemShut {NoStop}%
\bibitem [{\citenamefont {Kambersk{\`y}}(2007)}]{kambersky2007spin}%
  \BibitemOpen
  \bibfield  {author} {\bibinfo {author} {\bibfnamefont {V.}~\bibnamefont
  {Kambersk{\`y}}},\ }\bibfield  {title} {\bibinfo {title} {Spin-orbital
  {G}ilbert damping in common magnetic metals},\ }\href@noop {} {\bibfield
  {journal} {\bibinfo  {journal} {Physical Review B}\ }\textbf {\bibinfo
  {volume} {76}},\ \bibinfo {pages} {134416} (\bibinfo {year}
  {2007})}\BibitemShut {NoStop}%
\bibitem [{\citenamefont {Kune{\v{s}}}\ and\ \citenamefont
  {Kambersk{\`y}}(2002)}]{kunevs2002first}%
  \BibitemOpen
  \bibfield  {author} {\bibinfo {author} {\bibfnamefont {J.}~\bibnamefont
  {Kune{\v{s}}}}\ and\ \bibinfo {author} {\bibfnamefont {V.}~\bibnamefont
  {Kambersk{\`y}}},\ }\bibfield  {title} {\bibinfo {title} {{F}irst-principles
  investigation of the damping of fast magnetization precession in
  ferromagnetic 3 d metals},\ }\href@noop {} {\bibfield  {journal} {\bibinfo
  {journal} {Physical Review B}\ }\textbf {\bibinfo {volume} {65}},\ \bibinfo
  {pages} {212411} (\bibinfo {year} {2002})}\BibitemShut {NoStop}%
\bibitem [{\citenamefont {Kambersk{\`y}}(1984)}]{kambersky1984fmr}%
  \BibitemOpen
  \bibfield  {author} {\bibinfo {author} {\bibfnamefont {V.}~\bibnamefont
  {Kambersk{\`y}}},\ }\bibfield  {title} {\bibinfo {title} {{FMR} linewidth and
  disorder in metals},\ }\href@noop {} {\bibfield  {journal} {\bibinfo
  {journal} {Czechoslovak Journal of Physics B}\ }\textbf {\bibinfo {volume}
  {34}},\ \bibinfo {pages} {1111} (\bibinfo {year} {1984})}\BibitemShut
  {NoStop}%
\bibitem [{\citenamefont {Kambersk{\`y}}(1970)}]{KB}%
  \BibitemOpen
  \bibfield  {author} {\bibinfo {author} {\bibfnamefont {V.}~\bibnamefont
  {Kambersk{\`y}}},\ }\bibfield  {title} {\bibinfo {title} {On the
  {L}andau--{L}ifshitz relaxation in ferromagnetic metals},\ }\href@noop {}
  {\bibfield  {journal} {\bibinfo  {journal} {Canadian Journal of Physics}\
  }\textbf {\bibinfo {volume} {48}},\ \bibinfo {pages} {2906} (\bibinfo {year}
  {1970})}\BibitemShut {NoStop}%
\bibitem [{\citenamefont {Kambersk{\`y}}(1976)}]{TT}%
  \BibitemOpen
  \bibfield  {author} {\bibinfo {author} {\bibfnamefont {V.}~\bibnamefont
  {Kambersk{\`y}}},\ }\bibfield  {title} {\bibinfo {title} {On ferromagnetic
  resonance damping in metals},\ }\href@noop {} {\bibfield  {journal} {\bibinfo
   {journal} {Czechoslovak Journal of Physics B}\ }\textbf {\bibinfo {volume}
  {26}},\ \bibinfo {pages} {1366} (\bibinfo {year} {1976})}\BibitemShut
  {NoStop}%
\bibitem [{\citenamefont {Liechtenstein}\ \emph {et~al.}(1987)\citenamefont
  {Liechtenstein}, \citenamefont {Katsnelson}, \citenamefont {Antropov},\ and\
  \citenamefont {Gubanov}}]{liechtenstein1987local}%
  \BibitemOpen
  \bibfield  {author} {\bibinfo {author} {\bibfnamefont {A.~I.}\ \bibnamefont
  {Liechtenstein}}, \bibinfo {author} {\bibfnamefont {M.}~\bibnamefont
  {Katsnelson}}, \bibinfo {author} {\bibfnamefont {V.}~\bibnamefont
  {Antropov}},\ and\ \bibinfo {author} {\bibfnamefont {V.}~\bibnamefont
  {Gubanov}},\ }\bibfield  {title} {\bibinfo {title} {Local spin density
  functional approach to the theory of exchange interactions in ferromagnetic
  metals and alloys},\ }\href@noop {} {\bibfield  {journal} {\bibinfo
  {journal} {Journal of Magnetism and Magnetic Materials}\ }\textbf {\bibinfo
  {volume} {67}},\ \bibinfo {pages} {65} (\bibinfo {year} {1987})}\BibitemShut
  {NoStop}%
\bibitem [{\citenamefont {Papaconstantopoulos}\ and\ \citenamefont
  {Mehl}(2003)}]{papaconstantopoulos2003slater}%
  \BibitemOpen
  \bibfield  {author} {\bibinfo {author} {\bibfnamefont {D.}~\bibnamefont
  {Papaconstantopoulos}}\ and\ \bibinfo {author} {\bibfnamefont
  {M.}~\bibnamefont {Mehl}},\ }\bibfield  {title} {\bibinfo {title} {The
  {S}later-{K}oster tight-binding method: a computationally efficient and
  accurate approach},\ }\href@noop {} {\bibfield  {journal} {\bibinfo
  {journal} {Journal of Physics: Condensed Matter}\ }\textbf {\bibinfo {volume}
  {15}},\ \bibinfo {pages} {R413} (\bibinfo {year} {2003})}\BibitemShut
  {NoStop}%
\bibitem [{\citenamefont {Marzari}\ \emph {et~al.}(2012)\citenamefont
  {Marzari}, \citenamefont {Mostofi}, \citenamefont {Yates}, \citenamefont
  {Souza},\ and\ \citenamefont {Vanderbilt}}]{marzari2012maximally}%
  \BibitemOpen
  \bibfield  {author} {\bibinfo {author} {\bibfnamefont {N.}~\bibnamefont
  {Marzari}}, \bibinfo {author} {\bibfnamefont {A.~A.}\ \bibnamefont
  {Mostofi}}, \bibinfo {author} {\bibfnamefont {J.~R.}\ \bibnamefont {Yates}},
  \bibinfo {author} {\bibfnamefont {I.}~\bibnamefont {Souza}},\ and\ \bibinfo
  {author} {\bibfnamefont {D.}~\bibnamefont {Vanderbilt}},\ }\bibfield  {title}
  {\bibinfo {title} {Maximally localized {W}annier functions: {T}heory and
  applications},\ }\href@noop {} {\bibfield  {journal} {\bibinfo  {journal}
  {Reviews of Modern Physics}\ }\textbf {\bibinfo {volume} {84}},\ \bibinfo
  {pages} {1419} (\bibinfo {year} {2012})}\BibitemShut {NoStop}%
\bibitem [{\citenamefont {Pizzi}\ \emph {et~al.}(2020)\citenamefont {Pizzi},
  \citenamefont {Vitale}, \citenamefont {Arita}, \citenamefont {Bl{\"u}gel},
  \citenamefont {Freimuth}, \citenamefont {G{\'e}ranton}, \citenamefont
  {Gibertini}, \citenamefont {Gresch}, \citenamefont {Johnson}, \citenamefont
  {Koretsune} \emph {et~al.}}]{pizzi2020wannier90}%
  \BibitemOpen
  \bibfield  {author} {\bibinfo {author} {\bibfnamefont {G.}~\bibnamefont
  {Pizzi}}, \bibinfo {author} {\bibfnamefont {V.}~\bibnamefont {Vitale}},
  \bibinfo {author} {\bibfnamefont {R.}~\bibnamefont {Arita}}, \bibinfo
  {author} {\bibfnamefont {S.}~\bibnamefont {Bl{\"u}gel}}, \bibinfo {author}
  {\bibfnamefont {F.}~\bibnamefont {Freimuth}}, \bibinfo {author}
  {\bibfnamefont {G.}~\bibnamefont {G{\'e}ranton}}, \bibinfo {author}
  {\bibfnamefont {M.}~\bibnamefont {Gibertini}}, \bibinfo {author}
  {\bibfnamefont {D.}~\bibnamefont {Gresch}}, \bibinfo {author} {\bibfnamefont
  {C.}~\bibnamefont {Johnson}}, \bibinfo {author} {\bibfnamefont
  {T.}~\bibnamefont {Koretsune}}, \emph {et~al.},\ }\bibfield  {title}
  {\bibinfo {title} {Wannier90 as a community code: new features and
  applications},\ }\href@noop {} {\bibfield  {journal} {\bibinfo  {journal}
  {Journal of Physics: Condensed Matter}\ }\textbf {\bibinfo {volume} {32}},\
  \bibinfo {pages} {165902} (\bibinfo {year} {2020})}\BibitemShut {NoStop}%
\bibitem [{\citenamefont {Souza}\ \emph {et~al.}(2001)\citenamefont {Souza},
  \citenamefont {Marzari},\ and\ \citenamefont
  {Vanderbilt}}]{PhysRevB.65.035109}%
  \BibitemOpen
  \bibfield  {author} {\bibinfo {author} {\bibfnamefont {I.}~\bibnamefont
  {Souza}}, \bibinfo {author} {\bibfnamefont {N.}~\bibnamefont {Marzari}},\
  and\ \bibinfo {author} {\bibfnamefont {D.}~\bibnamefont {Vanderbilt}},\
  }\bibfield  {title} {\bibinfo {title} {Maximally localized {W}annier
  functions for entangled energy bands},\ }\href
  {https://doi.org/10.1103/PhysRevB.65.035109} {\bibfield  {journal} {\bibinfo
  {journal} {Phys. Rev. B}\ }\textbf {\bibinfo {volume} {65}},\ \bibinfo
  {pages} {035109} (\bibinfo {year} {2001})}\BibitemShut {NoStop}%
\bibitem [{\citenamefont {Roychoudhury}\ and\ \citenamefont
  {Sanvito}(2017)}]{roychoudhury2017spin}%
  \BibitemOpen
  \bibfield  {author} {\bibinfo {author} {\bibfnamefont {S.}~\bibnamefont
  {Roychoudhury}}\ and\ \bibinfo {author} {\bibfnamefont {S.}~\bibnamefont
  {Sanvito}},\ }\bibfield  {title} {\bibinfo {title} {Spin-orbit {H}amiltonian
  for organic crystals from first-principles electronic structure and {W}annier
  functions},\ }\href@noop {} {\bibfield  {journal} {\bibinfo  {journal}
  {Physical Review B}\ }\textbf {\bibinfo {volume} {95}},\ \bibinfo {pages}
  {085126} (\bibinfo {year} {2017})}\BibitemShut {NoStop}%
\bibitem [{\citenamefont {Mahfouzi}\ \emph {et~al.}(2017)\citenamefont
  {Mahfouzi}, \citenamefont {Kim},\ and\ \citenamefont
  {Kioussis}}]{PhysRevB.96.214421}%
  \BibitemOpen
  \bibfield  {author} {\bibinfo {author} {\bibfnamefont {F.}~\bibnamefont
  {Mahfouzi}}, \bibinfo {author} {\bibfnamefont {J.}~\bibnamefont {Kim}},\ and\
  \bibinfo {author} {\bibfnamefont {N.}~\bibnamefont {Kioussis}},\ }\bibfield
  {title} {\bibinfo {title} {Intrinsic damping phenomena from quantum to
  classical magnets: {A}n ab initio study of {G}ilbert damping in a pt/co
  bilayer},\ }\href {https://doi.org/10.1103/PhysRevB.96.214421} {\bibfield
  {journal} {\bibinfo  {journal} {Phys. Rev. B}\ }\textbf {\bibinfo {volume}
  {96}},\ \bibinfo {pages} {214421} (\bibinfo {year} {2017})}\BibitemShut
  {NoStop}%
\bibitem [{\citenamefont {Giannozzi}\ \emph {et~al.}(2009)\citenamefont
  {Giannozzi}, \citenamefont {Baroni}, \citenamefont {Bonini}, \citenamefont
  {Calandra}, \citenamefont {Car}, \citenamefont {Cavazzoni}, \citenamefont
  {Ceresoli}, \citenamefont {Chiarotti}, \citenamefont {Cococcioni},
  \citenamefont {Dabo} \emph {et~al.}}]{giannozzi2009quantum}%
  \BibitemOpen
  \bibfield  {author} {\bibinfo {author} {\bibfnamefont {P.}~\bibnamefont
  {Giannozzi}}, \bibinfo {author} {\bibfnamefont {S.}~\bibnamefont {Baroni}},
  \bibinfo {author} {\bibfnamefont {N.}~\bibnamefont {Bonini}}, \bibinfo
  {author} {\bibfnamefont {M.}~\bibnamefont {Calandra}}, \bibinfo {author}
  {\bibfnamefont {R.}~\bibnamefont {Car}}, \bibinfo {author} {\bibfnamefont
  {C.}~\bibnamefont {Cavazzoni}}, \bibinfo {author} {\bibfnamefont
  {D.}~\bibnamefont {Ceresoli}}, \bibinfo {author} {\bibfnamefont {G.~L.}\
  \bibnamefont {Chiarotti}}, \bibinfo {author} {\bibfnamefont {M.}~\bibnamefont
  {Cococcioni}}, \bibinfo {author} {\bibfnamefont {I.}~\bibnamefont {Dabo}},
  \emph {et~al.},\ }\bibfield  {title} {\bibinfo {title} {{QUANTUM ESPRESSO}: a
  modular and open-source software project for quantum simulations of
  materials},\ }\href@noop {} {\bibfield  {journal} {\bibinfo  {journal}
  {Journal of Physics: Condensed Matter}\ }\textbf {\bibinfo {volume} {21}},\
  \bibinfo {pages} {395502} (\bibinfo {year} {2009})}\BibitemShut {NoStop}%
\bibitem [{\citenamefont {Giannozzi}\ \emph {et~al.}(2017)\citenamefont
  {Giannozzi}, \citenamefont {Andreussi}, \citenamefont {Brumme}, \citenamefont
  {Bunau}, \citenamefont {Nardelli}, \citenamefont {Calandra}, \citenamefont
  {Car}, \citenamefont {Cavazzoni}, \citenamefont {Ceresoli}, \citenamefont
  {Cococcioni} \emph {et~al.}}]{giannozzi2017advanced}%
  \BibitemOpen
  \bibfield  {author} {\bibinfo {author} {\bibfnamefont {P.}~\bibnamefont
  {Giannozzi}}, \bibinfo {author} {\bibfnamefont {O.}~\bibnamefont
  {Andreussi}}, \bibinfo {author} {\bibfnamefont {T.}~\bibnamefont {Brumme}},
  \bibinfo {author} {\bibfnamefont {O.}~\bibnamefont {Bunau}}, \bibinfo
  {author} {\bibfnamefont {M.~B.}\ \bibnamefont {Nardelli}}, \bibinfo {author}
  {\bibfnamefont {M.}~\bibnamefont {Calandra}}, \bibinfo {author}
  {\bibfnamefont {R.}~\bibnamefont {Car}}, \bibinfo {author} {\bibfnamefont
  {C.}~\bibnamefont {Cavazzoni}}, \bibinfo {author} {\bibfnamefont
  {D.}~\bibnamefont {Ceresoli}}, \bibinfo {author} {\bibfnamefont
  {M.}~\bibnamefont {Cococcioni}}, \emph {et~al.},\ }\bibfield  {title}
  {\bibinfo {title} {Advanced capabilities for materials modelling with
  {Q}uantum {ESPRESSO}},\ }\href@noop {} {\bibfield  {journal} {\bibinfo
  {journal} {Journal of Physics: Condensed Matter}\ }\textbf {\bibinfo {volume}
  {29}},\ \bibinfo {pages} {465901} (\bibinfo {year} {2017})}\BibitemShut
  {NoStop}%
\bibitem [{\citenamefont {F{\"a}hnle}\ \emph {et~al.}(2013)\citenamefont
  {F{\"a}hnle}, \citenamefont {Steiauf},\ and\ \citenamefont
  {Illg}}]{fahnle2013erratum}%
  \BibitemOpen
  \bibfield  {author} {\bibinfo {author} {\bibfnamefont {M.}~\bibnamefont
  {F{\"a}hnle}}, \bibinfo {author} {\bibfnamefont {D.}~\bibnamefont
  {Steiauf}},\ and\ \bibinfo {author} {\bibfnamefont {C.}~\bibnamefont
  {Illg}},\ }\bibfield  {title} {\bibinfo {title} {Erratum: {G}eneralized
  {G}ilbert equation including inertial damping: {D}erivation from an extended
  breathing {F}ermi surface model [{P}hys. {R}ev. {B} 84, 172403 (2011)]},\
  }\href@noop {} {\bibfield  {journal} {\bibinfo  {journal} {Physical Review
  B}\ }\textbf {\bibinfo {volume} {88}},\ \bibinfo {pages} {219905} (\bibinfo
  {year} {2013})}\BibitemShut {NoStop}%
\bibitem [{\citenamefont {Wang}\ \emph {et~al.}(2022)\citenamefont {Wang},
  \citenamefont {Li}, \citenamefont {Wen}, \citenamefont {Cheng}, \citenamefont
  {Yin}, \citenamefont {Liu}, \citenamefont {Li},\ and\ \citenamefont
  {He}}]{wang2022two}%
  \BibitemOpen
  \bibfield  {author} {\bibinfo {author} {\bibfnamefont {H.}~\bibnamefont
  {Wang}}, \bibinfo {author} {\bibfnamefont {X.}~\bibnamefont {Li}}, \bibinfo
  {author} {\bibfnamefont {Y.}~\bibnamefont {Wen}}, \bibinfo {author}
  {\bibfnamefont {R.}~\bibnamefont {Cheng}}, \bibinfo {author} {\bibfnamefont
  {L.}~\bibnamefont {Yin}}, \bibinfo {author} {\bibfnamefont {C.}~\bibnamefont
  {Liu}}, \bibinfo {author} {\bibfnamefont {Z.}~\bibnamefont {Li}},\ and\
  \bibinfo {author} {\bibfnamefont {J.}~\bibnamefont {He}},\ }\bibfield
  {title} {\bibinfo {title} {Two-dimensional ferromagnetic materials: From
  materials to devices},\ }\href@noop {} {\bibfield  {journal} {\bibinfo
  {journal} {Applied Physics Letters}\ }\textbf {\bibinfo {volume} {121}}
  (\bibinfo {year} {2022})}\BibitemShut {NoStop}%
\end{thebibliography}%

\end{document}